\documentclass[graybox]{svmult}

\usepackage{type1cm}        % activate if the above 3 fonts are
\usepackage{makeidx}         % allows index generation
\usepackage{graphicx}        % standard LaTeX graphics tool
\usepackage{multicol}        % used for the two-column index
\usepackage[bottom]{footmisc}% places footnotes at page bottom

\usepackage{newtxtext}       % 
\usepackage[varvw]{newtxmath}       % selects Times Roman as basic font

\usepackage[utf8]{inputenc}
\usepackage[english]{babel}
\usepackage{amsmath,amsthm,amsfonts}
\usepackage{multirow}
\usepackage{tikz}
\usepackage{verbatim}
\usepackage{todonotes}
\usepackage{footmisc}
\usepackage{tocloft} % fixes TOC entry issues
\usepackage{isotope}
\usepackage{braket}
\usepackage{siunitx}
\DeclareSIUnit\torr{Torr} % to declare a new unit
\DeclareSIUnit\gauss{G}

\usepackage{bm}
\usepackage{braket}
\usepackage{microtype}
\usepackage{dsfont}
\DeclareMathOperator{\tr}{\rm{tr}}
\DeclareMathOperator{\M}{\mathcal{M}}

\usepackage{hyperref}
\hypersetup{
    colorlinks=true,
    linkcolor=blue,
    filecolor=blue,      
    urlcolor=cyan,
    citecolor=blue,
    pdftitle={Progress in quantum metrology and applications for optical atomic clocks},
    pdfpagemode=FullScreen,
    }

\makeindex             % used for the subject index
\begin{document}

\title*{Progress in quantum metrology and applications for optical atomic clocks}

\author{Raphael Kaubruegger \orcidID{0000-0003-0846-2333} and \\Adam M. Kaufman \orcidID{0000-0003-4956-5814}}

\institute{
Raphael Kaubruegger \at Department of Physics, University of Colorado, Boulder, CO 80309, USA  and \at JILA, University of Colorado and National Institute of Standards and Technology, Boulder, Colorado 80309, USA \email{raphael.kaubruegger@colorado.edu}
\and 
Adam M. Kaufman \at Department of Physics, University of Colorado, Boulder, CO 80309, USA  and \at JILA, University of Colorado and National Institute of Standards and Technology, Boulder, Colorado 80309, USA, \email{adam.kaufman@colorado.edu}
}

\maketitle

\abstract{Quantum entanglement offers powerful opportunities for enhancing measurement sensitivity beyond classical limits, with optical atomic clocks serving as a leading platform for such advances. This chapter introduces the principles of entanglement-enhanced quantum metrology and explores their applications to timekeeping. We review the theoretical framework of quantum phase estimation, comparing frequentist and Bayesian approaches, and discuss paradigmatic entangled states such as spin-squeezed and GHZ states. Particular emphasis is placed on the challenges posed by decoherence, which constrain the practical advantages that can be realized in large-scale devices. The discussion then turns to frequency estimation in atomic clocks, highlighting how experimental constraints shape the translation of abstract quantum limits into real performance gains. Finally, we outline emerging directions of contemporary quantum metrology. Together, these developments underscore the increasingly close interplay between quantum information processing and precision metrology.
}

\section{Introduction}

With few exceptions, quantum entanglement is the foundational resource of quantum-enhanced applications, including computing, metrology, networking, and simulation. In the context of metrology, entangled states can enhance signal-to-noise in resolving a parameter --- a magnetic field, a laser detuning, or a frequency difference --- while keeping other sensor properties --- particle number, total averaging time, etc. --- fixed. In this sense, entanglement serves as an axis to explore improvements in measurement science. 

The use of this axis is relevant when it becomes challenging to scale classical resources, like photon or atom number, or, when application-specific considerations limit achievable particle number, e.g. applications demanding high spatial resolution and therefore small form factor. In LIGO, the injection of non-classical squeezed light has improved the overall signal-to-noise, demonstrably increasing the radius of observation and thereby the frequency of observations of gravitational-wave events~\cite{caves1981quantum, grote2013first,tse2019quantum}. Remarkably, the use of squeezed light negotiates a fine balance between suppressing quantum projection noise from finite photon number and radiation-pressure noise from photon recoils, across the relevant sensitivity bandwidth of the kilometer-scale interferometer~\cite{ganapathy2023broadband}.

A key challenge in exploiting entangled resources lies both in their generation and in their fragility once prepared, requiring a demanding combination of precise control and isolation. In the context of sensing, this challenge is compounded by the fact that the goal is not only to make entangled states that are robust but also those that remain sensitive to the signals of interest. 

Spurred by the promise of future quantum computing applications, the progress in realizing large-scale programmable quantum systems of neutral atoms, ions and superconducting qubits has been simply stunning. These technological advances also present a unique opportunity for quantum sensing, as many of the underlying hardware platforms developed for quantum computing simultaneously serve as leading architectures for state-of-the-art quantum sensors~\cite{ludlow2015optical, ye2024essay}. In particular, trapped neutral atoms and ions are the basis of the best clocks in the world, and find application in many areas beyond time-keeping, including sensing of forces, accelerations, and magnetic and electric fields. Already since the early days of trapped-ion clocks, the translation of ideas from quantum information processing to spectroscopy—so-called quantum logic spectroscopy \cite{schmidt2005spectroscopy}—has been transformative. It has enabled access to ions such as aluminum, which form the basis of world-leading clocks but are otherwise challenging to control~\cite{ brewer2019al+, marshall2025high}. This same technique is now being used to realize new precision measurement directions for highly-charged ions, which offer a new set of directions for precision measurement and fundamental physics~\cite{kozlov2018highly,king2022optical}. 

With neutral-atom and ion-based platforms now capable of executing increasingly sophisticated quantum algorithms~\cite{bluvstein2024logical, ryan2024high}, the potential for implementing complex sensing protocols has expanded accordingly. In these notes, we provide an overview of several key themes within this rapidly evolving area of research. As this manuscript is primarily intended as an introduction, it does not claim to provide a comprehensive account of quantum metrology. For a more complete treatment of the subject, we refer the interested reader to existing review articles that capture the field as a whole in more detail ~\cite{demkowicz2015quantum, degen2017quantum, pezze2018quantum, ye2024essay, huang2024entanglement, montenegro2025quantum, kielinski2025bayesian}.

In Sect.~\ref{sec:QPE}, we provide an overview of quantum phase estimation, that is, the task of quantifying a phase that encodes a physical parameter of interest, such as a magnetic field or laser detuning. We then in Sect.~\ref{sec:frequentist} describe the so-called ``frequentist" approach to quantum phase estimation, which refers to the idealized problem of sensing a \emph{fixed}, unknown parameter a large number of times. This approach leads to a number of key limits on the achievable performance of quantum phase estimation. In this context, we discuss paradigmatic  quantum states that are used for entanglement-enhanced sensing applications in Sect.~\ref{sec:QuantumStates}.

While the frequentist approach simplifies analysis of the sensing problem, it omits the critical detail that one is often sensing a time-dependent parameter a few times, rather than a fixed parameter many times. How does this change the sensing problem theoretically, and what are the implied changes for how we should build experiments for quantum sensing?  
In Sect.~\ref{sec:Bayesian}, we introduce the framework of Bayesian quantum phase estimation, which addresses many of the aforementioned questions. This formalism facilitates the identification of quantum states and measurement strategies that enable high-precision phase estimation over a wide range of phase values.

In most cases, however, the primary quantity of interest is not the phase itself, but an underlying frequency that manifests as a phase in the quantum system after the atoms have interrogated the frequency for a fixed duration. In Sect.~\ref{sec:frequencyEstimation}, we examine the distinct challenges that arise in the context of frequency estimation and illustrate these using the example of an optical atomic clock.

In Sect.~\ref{sec:current}, we take a broader perspective and highlight selected ongoing experimental and theoretical efforts in the field of quantum metrology. While the experimental discussion primarily focuses on optical atomic clocks, the theoretical developments discussed extend to a wide range of quantum sensing platforms. Throughout this section, we draw upon the concepts and insights introduced in the preceding sections.

\section{Quantum Phase estimation}
\label{sec:QPE}
\subsection{Quantum phase estimation sequence}
The objective of quantum phase estimation is to use a quantum sensor composed of multiple quantum probes to determine an unknown phase encoded onto the quantum probes. This phase can be encoded, for example, by interactions of the probes with an external magnetic or electric field. Furthermore, the sensitivity of such a quantum sensor can be enhanced by leveraging entanglement among the probes. In general, a quantum phase estimation protocol comprises four steps: state preparation, phase encoding, measurement, and phase estimation.

More specifically, we consider the paradigmatic example of an ensemble of $N$ atoms, each characterized by two internal states, $\ket{\uparrow}$ and $\ket{\downarrow}$. Operations on this ensemble can be expressed in terms of the Pauli operators $\sigma_{\alpha}^{(k)}$ which act on the $k$-th particle, where $\alpha=x,y,z$ refer to the three orthogonal spin projections.

Usually an interferometric sequence with spin-$\tfrac{1}{2}$ atoms begins with the preparation of a coherent spin state (CSS),
\begin{equation}
    \ket{{\rm CSS}}=e^{-i\pi/2 J_y}\ket{\downarrow}^{\otimes N}
    \label{eq:CSS}
\end{equation} where all the spins are aligned along the $x$-axis. Here, the collective spin operators are defined as  $J_{\alpha}=\sum_{k=1}^N\sigma_{\alpha}^{(k)}/2$ which act uniformly on all of the atoms. In general any pure product state where all atoms are aligned in the same direction are referred to as CSS's. An entangled quantum state can be generated by applying a unitary operation $\mathcal{U}_{\rm en}$, referred to as the entangler, to a CSS. The resulting state can be described by the density matrix
\begin{equation}
    \rho=\mathcal{U}_{\rm en}\ket{{\rm CSS}}\bra{{\rm CSS}}\mathcal{U}^{\dagger}_{\rm en}
    \label{eq:initial_state}
\end{equation}
Here, we consider the case of a pure initial state; however, the formulation can be readily extended to accommodate an initially mixed state. In Sect.~\ref{sec:QuantumStates}, we will present concrete examples of entangled states along with their corresponding entangling operations that lead to enhanced sensitivity in sensing an unknown phase.

Following the state preparation, the quantum sensor undergoes unitary evolution of the form
\begin{equation}
    U_{\phi} =e^{-i\phi J_z}
\end{equation}
which encodes the unknown phase $\phi$ to be estimated, and $J_z$ being the generator of the phase encoding.
The quantum system after the phase encoding is described by the density matrix 
\begin{equation}
\rho_{\phi} = U_{\phi}\,\rho\, U^{\dagger}_{\phi}. 
\end{equation}

The encoded phase can be accessed through a measurement process. Here, we consider a von Neumann measurement defined as
\begin{equation}
    \mathcal{M} = \sum_{\mu}\mu\ket{\mu}\bra{\mu}
    \label{eq:VonNeumannMeasurement}
\end{equation}
where $\mu$ represents the possible measurement outcomes, and the set of states $\{\ket{\mu}\}$ forms an orthonormal basis of the $N$-atom Hilbert space. Alternatively, one may consider more general measurements described by positive operator-valued measures (POVMs) \cite{nielsen2010quantum}. These extend the projective von Neumann measurements introduced in Eq.\eqref{eq:VonNeumannMeasurement} to a set of measurement operators $\{\Pi_i\}$ that sum to the identity matrix $
\sum\Pi_i=\mathds{1}$ identity but are not required to be mutually orthogonal. For unitarily encoded single-parameter estimation, however, there always exists an optimal von Neumann measurement that attains the same sensitivity as any POVM, as discussed in Sect.\ref{sec:BCRB}. We therefore focus on projective measurements in the following.

In each repetition of the interferometric sequence, the measurement outcome $\mu$ is drawn from the conditional probability distribution
\begin{equation}
p(\mu|\phi) = \braket{\mu|\rho_{\phi}|\mu}. 
\label{eq:conditional_prob}
\end{equation}
A common measurement strategy consists of performing projective measurements on each atom individually in the computational basis ${\ket{\uparrow}, \ket{\downarrow}}$, assuming that the quantum sensor allows for single-site–resolved readout. The corresponding measurement basis can be written as $\ket{\boldsymbol{\sigma}} = \ket{\sigma_1, \dots, \sigma_N}$ with $\sigma_k \in \{\uparrow, \downarrow\}$. To access spin components along the $x$- or $y$-axes instead of the $z$-axis, an additional collective spin rotation is applied after the phase-encoding stage and prior to the projective measurement in the computational basis. 

Beyond simple spin rotations, a more general strategy is to apply a unitary transformation $\mathcal{U}_{\mathrm{de}}$ prior to the measurement. This operation, referred to as the decoder, enables projective measurements in a potentially entangled basis $\ket{\mu_{\boldsymbol{\sigma}}} = \mathcal{U}_{\mathrm{de}}\ket{\boldsymbol{\sigma}}$. Employing such a generalized measurement basis can increase the sensor’s robustness to noise and extend the range of phase values over which the encoded signal can be accurately inferred, as elaborated below.

The final step of the protocol involves converting the measurement outcomes into estimates of the encoded phase using an estimator function $\varphi$. The estimator function refers to the mathematical rule or algorithm that maps measurement outcomes to estimates of the unknown phase. Such functions can be constructed either analytically or numerically, provided that the response of the quantum sensor to the encoded phase is well characterized. Alternatively, the estimator can be calibrated empirically by probing the sensor with known input phases. Below, we provide examples of commonly used estimator functions. We distinguish between two types of estimators: single-measurement estimators $\varphi(\mu)$, which assign a phase estimate based on a single measurement outcome $\mu$, and frequentist estimators $\varphi(\bm\mu)$, which assume that the measurement protocol can be repeated $r$-times for the same encoded phase. In the latter case, the estimator is based on a vector of random measurement outcomes, $\bm\mu=(\mu_1, \dots, \mu_r)$ and the phase estimate is determined by considering all recorded measurement outcomes.

Regardless of the specific estimator used, it is essential to define a figure of merit that quantifies the deviation between the estimated value and the true parameter value. A commonly employed metric is the mean squared error (MSE), given by
\begin{equation}
    \varepsilon = \sum_{\mu}\left(\phi-\varphi(\mu)\right)^2p(\mu|\phi),
    \label{eq:MSE}
\end{equation}
which represents the squared estimation error averaged over the probability distribution of $p(\mu|\phi)$. The MSE provides a comprehensive assessment of the performance of the estimator, as it incorporates both accuracy and precision of the quantum sensor. This becomes evident when the MSE is rewritten as
 \begin{equation}
     \varepsilon=\left(\phi-\overline{\varphi}\right)^2+\Delta^2_{\varphi}. 
 \end{equation}
 The first term, $\left(\phi-\overline{\varphi}\right)^2$, quantifies the accuracy of the estimator by measuring how far the average estimated value $\overline{\varphi}=\sum_{\mu} \varphi(\mu)\,p(\mu|\phi)$ 
 deviates from the true parameter value $\phi$. This deviation is referred to as the bias of the estimator.

The second term, $\Delta^2_{\varphi}$, characterizes the precision of the estimator, capturing the distribution of the estimated values across different measurements. It is given by the variance of the estimator, 
\begin{equation}
    \Delta^2_{\varphi}=\sum_{\mu}\left(\varphi(\mu)-\overline{\varphi}\right)^2\, p(\mu|\phi)
\end{equation}

Thus, the MSE provides a unified metric of an estimator’s sensitivity, incorporating both systematic estimation errors (bias) and statistical fluctuations (variance) in the phase estimation process. One might be tempted to conclude that an unbiased sensor is intrinsically an accurate sensor; however, this view is incomplete. In practice, a quantum sensor can be affected by additional systematic errors. For example, in the case of optical atomic clocks, such errors may arise from blackbody radiation, residual interatomic interactions that cannot be fully suppressed and lead to interaction-induced frequency shifts, or relativistic frequency shifts, among others ~\cite{ludlow2015optical}. To substantiate a claim of accuracy at a given level, these systematic effects must be thoroughly characterized and corrected to within the stated uncertainty.

\section{Frequentist quantum phase estimation }
\label{sec:frequentist}
\subsection{Unbiased estimators}

In the frequentist statistical framework, the phase to be estimated is treated as an unknown but fixed parameter. Under this assumption an experimenter is able to repeat the sensing protocol  multiple times under identical conditions. In the limit of a large number of repetitions, the existence of an unbiased estimator is guaranteed, thereby justifying the minimization of the estimator’s variance while neglecting the potential bias. A similar justification applies in scenarios involving multiple quantum sensors that have the same phase encoded in their quantum states, effectively yielding repeated independent measurements of the same underlying parameter.

\begin{figure}[b]
   \includegraphics[]{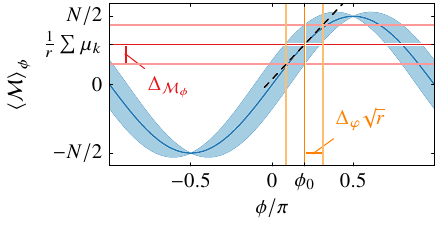}
   \caption{Functional dependence of the mean measurement outcome and its variance for a coherent spin state.
Shown is the expectation value of the observable $\braket{\mathcal{M}}{\phi}$ (blue line) and the corresponding measurement variance $\Delta^2{\mathcal{M}{\phi}} = \braket{\mathcal{M}^2}{\phi} - \braket{\mathcal{M}}{\phi}^2$ (blue shaded region), where $\mathcal{M} = J_y$ and the phase is encoded through the generator $J_z$. A measured sample mean $\frac{1}{r} \sum \mu_k$ can be converted into to a phase estimate $\varphi(\bm{\mu})$ by inverting the function $\braket{\mathcal{M}}_{\phi}$. In the limit of many measurement repetitions $r\gg 1$ the distribution of phase estimates is centered around the true encoded phase $\phi_0$ and with a variance $\Delta^2_{\varphi}$ that is proportional to the variance of the of the observable $\Delta^2_{\mathcal{M}_{\phi_0}}$ and the tangent of $\braket{\mathcal{M}}_{\phi}$ (black dashed line) at $\phi_0$.}
   \label{fig:SampleMeanEst}
\end{figure}

In this regime, a strategy is the sample mean estimator, which estimates the expectation value $\braket{\M}_{\phi}= \tr\left[\rho_{\phi}\M \right]$ from repeated measurement outcomes via the sample average $1/r\sum_{l=1}^r\mu_l$. If the functional dependence of the expectation value on the encoded phase is known, i.e., $\braket{\M}_{\phi}=f(\phi)$ —either through prior calibration of the sensor using deterministically encoded phases or through precise characterization of the system—this relation can be inverted to define the sample mean estimator (SME): 
\begin{equation}
    \varphi_{\rm SME}(\bm{\mu})=f^{-1}\left(\frac{1}{r}\sum_{l=1}^r\mu_l\right).
\end{equation}
In the limit of large $r$, the sample mean converges to the true expectation value, and the sample mean estimator becomes asymptotically unbiased within an interval where the function $f(\phi)$ is injective—i.e., where $\braket{\M}_{\phi}$ has no local extrema within the interval, except at the boundaries.

The variance of this estimator can be obtained via propagation of uncertainty (see Fig.~\ref{fig:SampleMeanEst}), yielding
\begin{equation}
    r\, \Delta^2_{\varphi}=\frac{\braket{\mathcal{M}^2}_{\phi} - \braket{\mathcal{M}}_{\phi}^2}{\left|\frac{\partial}{\partial \phi}\braket{\mathcal{M}}_{\phi}^2\right|^2}.
    \label{eq:SME_EstimatorVariance}
\end{equation}

The right-hand side of Eq.~\eqref{eq:SME_EstimatorVariance} is used as a benchmark to quantify the precision of a quantum sensor. This further highlights that there are two primary strategies to minimize the estimator variance for a given value of $\phi$: either by reducing the measurement variance, or by increasing the sensitivity of $\braket{\mathcal{M}}_{\phi}$ to changes in $\phi$, i.e., enhancing the magnitude of its derivative. In Sect.~\ref{sec:QuantumStates}, we discuss entangled quantum states that fulfill either of these criteria.

In practice, it is not known a priori how many measurements are required for the estimator variance to converge to the value predicted by the propagation of uncertainty. This is particularly relevant near phase values where $\braket{\mathcal{M}}_\phi$ has extrema, as more repetitions may be needed in these regions. 

\begin{figure}[b] 
   \centering
   \includegraphics[]{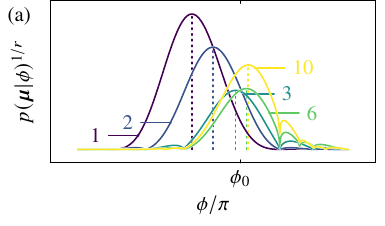}\\
   \includegraphics[]{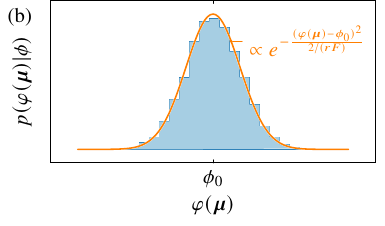}
   \caption{(a) Schematic of a likelihood function $p(\bm{\mu}|\phi)$ for $r= 1,\dots,10$ simulated measurement outcomes. As the number of measurements increases, the maximum of the likelihood function (dashed lines) converges, on average over many realizations of the simulated measurements, to the true encoded phase $\phi_0$ (b) Histogram of the distribution of simulated maximum likelihood estimates, compared to a Gaussian distribution centered at the true encoded phase value $\phi_0$ with standard deviation $\sqrt{1/(r\, F)}$, where $F$ is the Fisher information of the conditional probability distribution $p(\mu|\phi)$.}
   \label{fig:MLE}
\end{figure}

Since the sample mean estimator is not always optimal, frequentist approaches to quantum phase estimation often employ the maximum likelihood estimator (MLE). The MLE is defined as the value of the phase $\phi$ that maximizes the likelihood function $p(\bm\mu|\phi)=\prod_{k=1}^rp(\mu_k|\phi)$, i.e., the probability of obtaining the observed set of measurement outcomes $\boldsymbol{\mu}$ given the encoded phase is $\phi$:
\begin{equation}
    \varphi_{\rm MLE}(\bm \mu)= {\rm argmax}_{\phi}p(\bm \mu|\phi), 
    \label{eq:MLE}
\end{equation}

To build intuition for the concept of the maximum-likelihood function, consider a single atom for which the probability of measuring the spin in the excited state $\ket{\uparrow}$ is given by $p(\uparrow|\phi) = \frac{1+\sin(\phi)}{2}$, and correspondingly the probability of measuring it in the ground state $\ket{\downarrow}$ is $p(\downarrow|\phi) = \frac{1-\sin(\phi)}{2}$. Suppose that, after $r$ repeated measurements, the atom is found in the excited state $r_{\uparrow}$ times and in the ground state $r_{\downarrow}=r-r_{\uparrow}$ times. The corresponding likelihood function is then
$p(r_{\uparrow}, r_{\downarrow}|\phi) = \left(\frac{1+\sin(\phi)}{2}\right)^{r_{\uparrow}} \left(\frac{1-\sin(\phi)}{2}\right)^{r_{\downarrow}}$.
Here, we assume that the order of measurement outcomes is irrelevant, which follows from the initial assumption that the encoded phase $\phi$ remains constant over all repetitions of the measurement.

Figure~\ref{fig:MLE} shows a simulated MLE experiment, illustrating the property of asymptotic efficiency: in the limit of many repetitions, the estimator converges to the true phase value, provided the phase lies within the region of unambiguous estimation. This property is a primary reason for the widespread use of MLE. Asymptotic efficiency further implies that the distribution of phase estimates obtained from repeated trials is well approximated by a normal distribution with variance $1/(rF)$. Here, $F$ denotes the Fisher information, defined in detail below, which quantifies the sensitivity of the quantum state and measurement scheme to the parameter of interest. The Fisher information is associated with the conditional probability distribution $p(\mu|\phi)$, often referred to as the likelihood function, as it specifies the probability of obtaining outcome $\mu$ for a given encoded phase $\phi$. Consequently, the MLE saturates the Cramér–Rao bound (CRB) in the large-sample limit, achieving the minimal possible variance for any unbiased estimator, as we elaborate in the following section.

\subsection{Cramér-Rao bound}
The CRB ~\cite{frechet1943extension, rao1992information, cramer1999mathematical}
\begin{equation}
    r\, \Delta^2_{\varphi}\geq \frac{1}{F_{\rho_\phi, \ket{
    \mu
    }}}
    \label{eq:CRB}
\end{equation}
a central quantity in frequentist estimation theory states that the variance of an unbiased estimator is lower bounded by the FI of the probability distribution. The FI 
\begin{equation}
    F_{\rho_\phi,\ket{\mu}}=\sum_{\mu} \left(\frac{\partial}{\partial\phi}\log\left[p(\mu|\phi)\right]\right)^2p(\mu|\phi)
    \label{eq:FisherInformation}
\end{equation}
quantifies the amount of information that the probability distribution $p(\mu|\phi)$ carries about an unknown parameter $\phi$. It does so by combining the squared derivatives of the log-likelihood—capturing the sensitivity to infinitesimal changes in $\phi$—with weights given by the likelihood of observing each measurement outcome.

The FI—and thus the CRB—is independent of the particular choice of estimator. Specifically, the CRB represents the minimum variance attainable by any unbiased estimator. The MLE is guaranteed to asymptotically saturate this bound in the limit of large $r$. As such, the CRB serves as a valuable benchmark for evaluating the performance of suboptimal estimators and for quantifying the potential gains in precision achievable through improved estimation strategies.

In practice, it is often assumed that the number of measurements required for an estimator to become approximately unbiased and to approach the CRB is not prohibitively large. However, there exists no general method for predicting the number of measurements required for this convergence. Instead, simulated estimation experiments can be used to develop intuition about estimator performance in realistic settings. In Sect.~\ref{sec:QuantumStates}, we present sample results from such simulations, highlighting the subtlety that states with higher FI often require a larger number of measurements to realize their metrological advantage.

\subsection{Quantum Cramér-Rao bound}
An upper bound to the FI is obtained by maximizing over all possible quantum measurements. For single-parameter estimation von Neumann measurement are sufficient to saturate this bound ~\cite{helstrom1967minimum, personick1971application}. The maximal FI achievable for a given quantum state, optimized over all possible measurements, is known as the QFI ~\cite{braunstein1994statistical}, and the corresponding lower bound on the variance of any unbiased estimator is referred to as the quantum Cramér-Rao bound (QCRB) ~\cite{helstrom1967minimum}
\begin{equation}
    r\, \Delta^2_{\varphi}\geq \frac{1}{Q_{\rho_{\phi}}}.
    \label{eq:QCRB}
\end{equation}
For a pure quantum state, the quantum Fisher information (QFI) is given by four times the variance of the generator of the parameter encoding ~\cite{fujiwara1995quantum}. In the context discussed here, where the generator is $J_z$, the QFI is therefore
\begin{equation}
    Q_{\rho_{\phi}=\ket{\psi}\bra{\psi}}=4\left(\braket{\psi|J_z^2|\psi}-\braket{\psi|J_z|\psi}^2\right).
    \label{eq:QFI_pure}
\end{equation}

The previously introduced expression for the QFI only applies for pure initial states. To treat the general case of a parameter-dependent density matrix of the form $\rho_{\phi} = e^{-i\phi J_z} \rho_0\, e^{+i\phi J_z}$, the QFI can be conveniently expressed in terms of the symmetric logarithmic derivative (SLD), denoted by $L_{\phi}$. The SLD is defined as the self-adjoint operator satisfying the operator equation
\begin{equation}
    \frac{\partial}{\partial \phi}\rho_{\phi}=-i\left[J_z,\rho_{\phi}\right]=\frac{L_{\phi}\rho_{\phi}+\rho_{\phi}L_{\phi}}{2}. 
    \label{eq:SLD}
\end{equation}
The QFI can then be written in compact form ~\cite{yuen1973multiple, helstrom1974noncommuting, holevo2011probabilistic, braunstein1994statistical, braunstein1996generalized} as 
\begin{equation}
Q_{\rho_{\phi}}= \mathrm{Tr}\left[\rho_{\phi} L_{\phi}^2\right].
\end{equation}
For a unitary phase encoding, the SLD transforms as $L_{\phi} = e^{-i\phi J_z} L_0\, e^{+i\phi J_z}$, which shows that the QFI is independent of the actual value of the encoded phase $\phi$ and can therefore be evaluated entirely from the initial density matrix $\rho_0$.

To obtain an explicit form for the SLD, we consider the spectral decomposition of the initial density matrix,
$\rho_0 = \sum_k p_k\, \ket{\psi_k}\bra{\psi_k}$,
where ${\ket{\psi_k}}$ denotes an orthonormal states and $p_k$ are the probabilities that the respective states occur. The number of nonzero eigenvalues defines the rank of the density matrix, with pure states corresponding to the special case of rank one.

In practice the SLD can be obtained as
\begin{equation}
    L_{0}=2\sum_{k, l}\frac{-i\braket{\psi_{ k}|\left[J_z,\rho_{0}\right]|\psi_{ l}}}{p_k+p_l}\ket{\psi_{ k}}\bra{\psi_{l}}, 
\end{equation}
which also constitutes a measurement that saturates the QCRB. Using this form of the SLD one can represent the QFI as~\cite{liu2014quantum}  
\begin{eqnarray}
    Q_{\rho_{0}}= &\sum_{i}p_i\, Q_{\ket{\psi_{k}}\bra{\psi_{k}} } \notag \\
    &- \sum_{k\neq l}\frac{8 p_k p_l}{p_k + p_l}\left|\bra{\psi_{k}}J_z\ket{\psi_{l}}\right|^2.
    \label{eq:QFI}
    \end{eqnarray}
The expression in Eq.\eqref{eq:QFI} consists of a convex sum of the QFI of the pure states in the spectral decomposition of the density matrix, subtracted by strictly positive terms. This implies that the convex sum always serves as an upper bound to the QFI of the mixed state. It is worth noting that an alternative expression for the QFI exist; however, they are only well-defined when the density matrix has full rank. In contrast, the formulation in Eq.\eqref{eq:QFI} remains valid for density matrices of arbitrary rank.

An optimal measurement that saturates the QCRB can be implemented by performing a projective measurement in the eigenbasis of the SLD. A drawback of this approach is that both the SLD and the corresponding optimal measurement generally depend on the value of the unknown parameter. Consequently, the QCRB can only be saturated when sufficient prior knowledge of the parameter is available. This limitation can be overcome by adaptive measurement strategies, in which the measurement basis is iteratively updated based on the outcomes of earlier measurements~\cite{hayashi2005asymptotic}.

\subsection{Quantum limits on the estimator variance}
\label{sec:qfi_limits}
Having identified the QFI as a quantity that is already optimized over all unbiased estimators and all possible measurements, the remaining task is to determine which quantum states maximize the QFI. This corresponds to optimizing over all entangling unitaries $\mathcal{U}_{\rm en}$ which allow the preparation of any quantum state. We first consider the class of quantum states that do not share entanglement between atoms, such that the overall state can be written as a tensor product of single-atom density matrices, $\rho_{\rm sep} = \rho^{(1)} \otimes \rho^{(2)} \otimes \cdots \otimes \rho^{(N)}$.
In this case, the availability of $N$ uncorrelated atoms, which can be considered as independent sensors, is equivalent to repeating the same measurement $N$ times. Consequently, the estimator variance satisfies
\begin{equation}
r\Delta^2_{\varphi,\rm sep} \geq \frac{1}{Q_{\rho_{\rm sep}}} \geq \frac{1}{N},
\end{equation}
which is known as the standard quantum limit (SQL) of quantum phase estimation. The SQL illustrates an important property of the QFI, namely its additivity for separable subsystems. If the density matrix factorizes as $\rho_{A} \otimes \rho_{B}$, the corresponding QFI is additive:
\begin{equation}
    Q_{\rho_{A} \otimes \rho_{B}} = Q_{\rho_{A}} + Q_{\rho_{B}}.
\end{equation}

The central promise of quantum metrology is that the SQL can be surpassed by entangling the constituent atoms, prompting the question of what fundamental precision limit is imposed by quantum mechanics. Since the QFI of a mixed state is upper bounded by the convex sum of the QFIs of states that appear in the spectral decomposition of the density matrix, the maximal QFI is achieved by a pure state that maximizes the QFI in Eq.~\eqref{eq:QFI_pure}. Such a state can always be constructed as an equal superposition of the eigenstates corresponding to the maximal and minimal eigenvalues of the generator. Thus, for the generator $J_z$, the state that maximizes QFI is 
\begin{equation}
\ket{\psi_{\rm GHZ}} = \frac{\ket{\uparrow}^{\otimes N} + \ket{\downarrow}^{\otimes N}}{\sqrt{2}},
\label{eq:GHZ_state}
\end{equation}
which is commonly referred to as the GHZ state \cite{greenberger1989going} (see Sect.~\ref{sec:GHZ_state} for a more in depth discussion of the GHZ-state). The QFI of the GHZ state is $Q_{\ket{\psi_{\rm GHZ}}\bra{\psi_{\rm GHZ}}} = N^2$, which sets the fundamental lower bound on the estimator variance:
\begin{equation}
r\, \Delta^2_{\varphi} \geq  \frac{1}{N^2}.
\end{equation}
This bound is commonly referred to as the Heisenberg limit (HL) ~\cite{giovannetti2006quantum} of quantum phase estimation and cannot be surpassed by any entangled state composed of $N$ spin-$\frac{1}{2}$ atoms. The HL can be generalized to arbitrary generators of the phase encoding. In this case the HL is the squared difference between the eigenvalue extrema of the generator, i.e., $(\lambda_{\max} - \lambda_{\min})^2$, where $\lambda_{\max}$ and $\lambda_{\min}$ denote the largest and smallest eigenvalues of the generator, respectively. Consequently, if the phase is encoded via a different operator—for example, $J_z^2$ instead of $J_z$—the corresponding HL may have a different scaling, potentially exceeding the conventional $N^2$ scaling associated with phase encoding via $J_z$.

\subsection{Multipartite entanglement}
The HL, together with the additivity of the quantum Fisher information (QFI), establishes that the QFI, the classical Fisher information, and the inverse estimator variance of an unbiased estimator can serve as witnesses of multipartite entanglement~\cite{hyllus2012fisher, toth2012multipartite}. This can be understood by partitioning the $N$ atoms into $l = \left\lfloor N/k \right\rfloor$ subgroups of $k$ atoms each, along with a smaller group containing $N - k\,l$ atoms, assuming no entanglement between the different partitions,  where  $\left\lfloor N/k \right\rfloor$ denotes the greatest integer less than or equal to $N/k$. The state that maximizes the QFI under this constraint is one in which each partition is prepared in a GHZ state, yielding a total QFI of
$Q_{\rho_{k\text{-partite}}} = l\,k^2 + (N - k\,l)^2$.
Hence, any state whose QFI or inverse estimator variance exceeds $l\,k^2$ must contain either tensor products of ($k+1$)-atom GHZ states or larger entangled partitions. In other words, such a state must be at least ($k+1$)-partite entangled, making the QFI, FI, and inverse estimator variance valid witnesses of multipartite entanglement.

In general, a quantum state represented by a density matrix is said to possess $k$-partite entanglement if its QFI cannot be reproduced by partitioning the system into subsystems containing at most $k-1$ atoms each, under the assumption that every subsystem is prepared in a GHZ state, which maximizes the QFI for the subsystem. Equivalently, if reproducing the QFI of the full state requires at least one subsystem containing $k$ atoms, then the state must exhibit at least $k$-partite entanglement.

It is important to emphasize that the concept of multipartite entanglement is distinct from the commonly studied notion of bipartite entanglement, which is typically quantified using the von Neumann entropy of a reduced subsystem ~\cite{nielsen2010quantum}. For example, an $N$-atom GHZ state has maximal multipartite entanglement, yet contains only a limited amount of bipartite entanglement. In other words, the hierarchy of quantum states according to their bipartite entanglement content can differ substantially from that obtained when ranking them by their multipartite entanglement content. Nevertheless, both measures indicate the presence of genuine entanglement among the constituent atoms.

\section{Quantum states for sensing}
\label{sec:QuantumStates}
In this section we introduce three different families of states that are commonly discussed in the context and highlight some of their strengths and flaws, as well as the measurement required to achieve the QCRBs for the respective states. 

\subsection{Coherent spin states}

\begin{figure}[b] 
   \centering
   \includegraphics[]{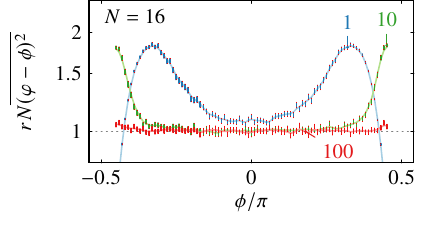}
   \caption{Comparison of the squared estimation errors for a coherent spin state composed of $N = 16$ atoms, obtained from $r = 1$, $10$, and $100$ randomly sampled measurement outcomes in the $y$-basis, averaged over $10^4$ simulated estimation experiments. For a sensor consisting of uncorrelated atoms and in the absence of additional noise, the estimation errors of the maximum-likelihood estimator and the sample-mean estimator coincide. The gray dotted line indicates the standard quantum limit.}
   \label{fig:CSS_MLE_SME}
\end{figure}

The reference point for any quantum metrological enhancement is the spin coherent state defined in Eq.~\eqref{eq:CSS}, which consists of N unentangled atoms and achieves, at best, the SQL. In the following, we consider the specific spin coherent state $\ket{\psi_{\rm CSS}} = e^{+i\pi/2 J_y} \ket{\downarrow}^{\otimes N}$, in which all spins are aligned along the $x$-axis. This state is characterized by the expectation values $\braket{J_x} = N/2$, $\braket{J_{y,z}} = 0$, and $\braket{J_{y,z}^2} = N/4$. From this, it follows that a measurement of $J_y$ yields an estimator variance of $r\, \Delta^2_{\varphi} = 1/N$, which saturates the SQL. 

A fundamental limitation of sensors based on CSSs with a measurement of $J_y$ arises from the periodicity of the expectation value $\braket{J_y}_{\phi} = \frac{N}{2}\sin(\phi)$. Because of this sinusoidal dependence, phase shifts exceeding $\pi/2$ cannot be distinguished from those smaller than $\pi/2$, restricting the unambiguous phase-estimation range to $\phi \in [-\pi/2, \pi/2]$. Consequently, such sensors provide unique phase estimates only within this interval. Besides this, the CRB is independent of the encoded phase and reaches the SQL for all values of $\phi$. Achieving the CRB, however, requires averaging over many measurement outcomes, and the number of repetitions needed depends on the encoded phase, as illustrated in Fig.~\ref{fig:CSS_MLE_SME}. For a sensor composed of $N=16$ atoms, a single measurement is insufficient to saturate the CRB. The number of measurements required to approach this bound varies with $\phi$ and  estimates near the edges of the unambiguous phase-estimation interval demand substantially more repetitions.

Moreover, for $r=1$ the squared estimation error falls below the CRB as $\phi \to \pm \pi/2$. This behavior arises because, at $\phi=\pm \pi/2$, the coherent state is an eigenstate of the measurement operator, such that measurement noise no longer contributes to the estimation error. At the same time, the MLE becomes increasingly biased in this regime, rendering the CRB inapplicable as a lower bound. Consequently, the observed estimation error can drop below the nominal CRB, and the same effect arises for larger values of $r$ as $\phi$ approaches $\phi=\pm \pi/2$ as well. 

\subsection{Spin squeezed states}
In a seminal work, Wineland et al. introduced the concept of spin squeezing ~\cite{wineland1992spin}. They demonstrated that, for a collective spin initially polarized along the $x$-axis, where phase information is typically extracted via a measurement of $J_y$, the estimator variance can be reduced by decreasing the variance $\Delta^2_{J_y} = \braket{J_y^2} - \braket{J_y}^2$ (which determines the numerator of the estimator variance) more rapidly than the squared mean spin $\braket{J_x}^2$ (which appears in the denominator). This reduction, however, comes at the expense of an increased variance $\Delta^2_{J_y}$, which can be understood in terms of the Heisenberg uncertainty relation for non-commuting spin observables 
\begin{equation}
\Delta_{J_y}^2 \Delta_{J_z}^2 \geq \braket{J_x}^2 / 4. 
\end{equation}
In other words, a state is considered a spin squeezed state (SSS) if the quantum fluctuations along an axis orthogonal to the mean spin direction are reduced, while the fluctuations along the third, mutually orthogonal direction are correspondingly increased. Wineland et al. quantified the enhancement in phase estimation precision as the ratio between the estimator variance of an entangled state, given by Eq.~\eqref{eq:SME_EstimatorVariance}, and the minimum estimator variance attainable with an unentangled state containing the same number of atoms. For the specific case of a $J_y$ measurement, this expression reduces to the Wineland spin-squeezing parameter,
\begin{equation}
\xi^2 = N \frac{\braket{J_y^2} - \braket{J_y}^2}{\braket{J_x}^2}. 
\end{equation}
For a comprehensive overview of related spin-squeezing parameters, the reader is referred to ~\cite{ma2011quantum}. In general, a ratio smaller than unity implies the presence of metrologically-useful entanglement, as it indicates a reduction of the estimator variance below the limit achievable with unentangled atoms. Consequently, a state is said to be spin-squeezed when $\xi^2 < 1$, and the observation of spin squeezing necessarily implies the presence of entanglement among the atoms.

Extreme SSSs, defined as the states that minimize the variance $\Delta^2 J_y$ for a given expectation value $\langle J_x\rangle$, can be obtained as the ground states of the Hamiltonian
~\cite{sorensen2001entanglement}
\begin{equation}
    H = \chi J_y^2+\omega J_x, 
    \label{eq:SqueezingParent}
\end{equation}
where the ratio $\omega/\chi$ serves as a control parameter to tune $\braket{J_x}$. For even $N$, the spin squeezing parameter is minimized in the limit $\omega/\chi \rightarrow 0$, where it approaches the fundamental bound on spin squeezing ~\cite{hillery1993interferometers, agarwal1994atomic}
\begin{equation}
    \xi^2\geq \frac{2}{N+2},
    \label{eq:squeezing_phase}
\end{equation}
which shows Heisenberg scaling.

In this limit, the ground state approaches the Dicke state $e^{-i\pi/2J_x}\ket{0}$, rotated such that it is an eigenstate of $J_y$ instead of $J_z$. Dicke states $\ket{M}$ are the unique states that are invariant under atom permutations. They satisfy the eigenvalue equations  $J_z\ket{M}=M\ket{M}$ and $(J_x^2+J_y^2+J_z^2)\ket{M}=N/2(M/2+1)\ket{M}$. 

While the state $e^{-i\pi/2J_x}\ket{0}$ is theoretically optimal in terms of spin squeezing, it suffer from a fundamental drawback: both the numerator and denominator in the squeezing parameter become vanishingly small, resulting in a finite value only in the idealized limit. In practice, any small amount of additional noise can dominate the vanishing variance in the numerator, and due to the small denominator, lead to a significant amplification of this noise, ultimately resulting in poor estimation performance. It is therefore advisable to avoid states that are close to being eigenstates of the measured observable.

Spin-squeezed states can be obtained not only as ground states of suitable Hamiltonians but also through dynamical preparation. A paradigmatic example is the quench of a CSS initially aligned along the $x$-axis under the so-called one-axis twisting (OAT) Hamiltonian ~\cite{kitagawa1993squeezed},
\begin{equation}
H_{\rm OAT} = \chi J_z^2.
\label{eq:H_OAT}
\end{equation}
For references to experimental realizations of this Hamiltonian, and for recent efforts to prepare SSSs using finite-range interactions, refer to Sect.\ref{sec:short_range_squeezing}.

Under a single quench of the OAT Hamiltonian, i.e.
$\ket{\psi_{\mathrm{OAT}}(t)}=e^{-i t H_{\mathrm{OAT}}}\,e^{-i\frac{\pi}{2} J_y}\ket{\downarrow}^{\otimes N}$,
the minimal Wineland spin-squeezing parameter scales as $\xi^2_{\mathrm{OAT}}\propto N^{-2/3}$ at an evolution time $\chi t \propto N^{-2/3}$ ~\cite{kitagawa1993squeezed}. For longer evolution times, the state becomes oversqueezed: its metrological advantage is no longer captured by the squeezing parameter $\xi^2$ and cannot be accessed through a simple $J_y$ measurement, even though both the FI and the QFI continue to increase. This additional, metrological gain can be quantified using nonlinear squeezing parameters~\cite{gessner2019metrological}.

\begin{figure}[b] 
   \centering
   \includegraphics[]{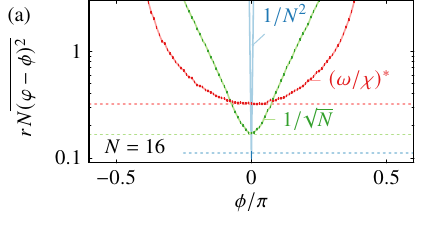}\\
   \includegraphics[]{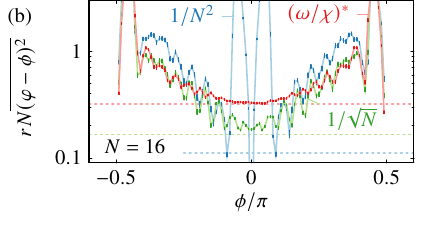}
   \caption{Comparison of the squared estimation errors for spin-squeezed states with $N = 16$ atoms, based on $r = 10$ measurement repetitions in the $y$-basis and averaged over $1000$ simulated estimation experiments for a sample mean estimator  (a) and a maximum likelihood estimator (b). The states correspond to the ground states of the spin-squeezing Hamiltonian [Eq.\eqref{eq:SqueezingParent}] for $\omega/\chi = 1 / N^2, 1 / \sqrt{N}, (\omega/\chi)^*$, where $(\omega/\chi)^*$ denotes the value at which $\Delta^2_{J_y}=\Delta^2_{J_x}$. Dashed lines indicate the phase-dependent sample mean estimator variance [Eq.\eqref{eq:squeezing_phase}] for each corresponding state. Dotted lines indicate the Cram\'er-Rao bound for each spin squeezed state state.
   }
   \label{fig:SSS}
\end{figure}

Spin-squeezed states highlight a fundamental distinction between the SME and the MLE. Specifically, the estimator variance of the SME for a $J_y$ measurement is given by
\begin{equation}
r\,\Delta_{\varphi}^2=\frac{\cos^2(\phi)\Delta_{J_y}^2+\sin^2(\phi)\Delta^2_{J_x}}{\cos^2(\phi)\braket{J_x}^2},
\label{eq:phase_dependent_squeezing}
\end{equation}
where the observable is evaluated in the rotated frame resulting from the phase encoding, i.e., $J_y \rightarrow e^{i\phi J_z} J_y e^{-i\phi J_z} = \cos(\phi)J_y - \sin(\phi)J_x$. This expression reveals an explicit dependence on the encoded phase $\phi$; the estimator variance attains the minimum value associated with the spin squeezing parameter only at $\phi = 0$, and increases for all other values. Notably, the stronger the spin squeezing, the more pronounced this increase in estimator variance becomes away from $\phi = 0$ ~\cite{andre2004stability, braverman2018impact}. This behavior can be understood from Eq.~\eqref{eq:phase_dependent_squeezing} as arising from an admixture of the variance along the mean spin direction. While the CSS is an eigenstate of $J_x$ and therefore has zero variance along this direction, this variance is increased as the state becomes increasingly squeezed.

Figure~\ref{fig:SSS} (a) compares SSSs with different degrees of squeezing. It clearly shows that the state with the strongest squeezing—characterized by a squeezing parameter approaching $2/(N+2)$ has a significantly narrower interval over which the encoded parameter can be reliably estimated. Although using a less strongly squeezed state increases the estimation range, it remains narrower than that achievable with a CSS.

In contrast, the FI for a SSS measured in the $y$-basis is independent of the encoded phase $\phi$, which might suggest that the estimator variance of a MLE is likewise independent of $\phi$. However, in practice, the number of measurements required to approach the CRB has a pronounced dependence on the encoded phase, as illustrated in Fig.~\ref{fig:SSS} (b). It is important to emphasize that for finite sample sizes, the MLE is generally biased and can have averaged estimation errors that lie below the CRB. In the asymptotic limit $r \to \infty$, the MLE ultimately saturates the CRB across all phase values, whereas the estimation error associated with the SME remains largely unaffected, as it already approaches the lower sensitivity bound predicted by Eq.~\eqref{eq:phase_dependent_squeezing} for $r = 10$.

\subsection{GHZ-state}
\label{sec:GHZ_state}

\begin{figure}[b] 
   \centering
   \includegraphics[]{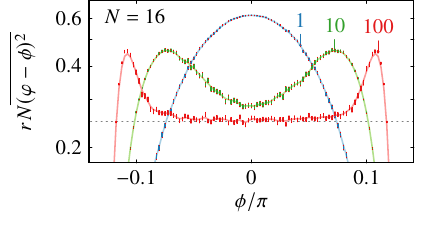}
   \caption{Comparison of the squared estimation errors obtained using a GHZ state with $N = 16$ atoms, evaluated for $r = 1, 10, 100$ measurement outcomes in the basis of $\Pi_x$ and averaged over 10000 simulated estimation experiments. In the absence of noise, the sample mean estimator and the maximum likelihood estimator show the same performance. Note that the scale of the horizontal axis is compressed by a factor of $N$ relative to Figs.~\ref{fig:CSS_MLE_SME}, and \ref{fig:SSS}. The gray dashed line denotes the Heisenberg limit. In this example, the GHZ state has been rotated to ensure symmetric sensitivity around $\phi = 0$.}
   \label{fig:GHZ_SME}
\end{figure}

Another paradigmatic state in quantum metrology is the GHZ state, originally introduced by Greenberger, Horne, and Zeilinger ~\cite{greenberger1989going} as an example of quantum nonlocality, yielding predictions incompatible with any classical deterministic local theory. The GHZ state is defined as the equal superposition of all atoms in the ground state and all atoms in the excited state:
\begin{equation}
\ket{\psi_{\rm GHZ}} = \frac{1}{\sqrt{2}} \left( \ket{\uparrow}^{\otimes N} + \ket{\downarrow}^{\otimes N} \right).
\label{eq:GHZ_state2}
\end{equation}
Unlike SSSs, the metrological advantage of GHZ states does not arise from a reduction in measurement noise, but rather from an enhanced sensitivity to the encoded phase. Specifically, GHZ states accumulate phase at a rate $N$ times faster than that of a single spin-$\tfrac{1}{2}$ particle. Under unitary evolution $U_{\phi}$, the GHZ state evolves as $U_{\phi} \ket{\psi_{\rm GHZ}} \propto e^{+i\phi \frac{N}{2}}\ket{\uparrow}^{\otimes N} + e^{-i\phi \frac{N}{2}}\ket{\downarrow}^{\otimes N}$,  illustrating the $N$-fold amplification of the acquired phase.

However, the phase amplification provided by GHZ states cannot be exploited using standard single-body observables such as $J_y$, as these operators are sensitive only to coherences between states that differ by a single spin flip. In contrast, the GHZ state has coherences exclusively between states that differ by $N$ spin flips. As a result, observables capable of probing this structure must be tailored accordingly. One such observable, proposed in Ref.~\cite{bollinger1996optimal}, is the $N$-body operator $\Pi_x = \prod_{k=1}^N \sigma_x^{(k)}$, commonly referred to as the $x$-parity operator.

To verify that the GHZ state and a parity measurement saturate the HL, one can evaluate the relevant expectation values: $\braket{\Pi_x}_{\phi} = \cos(N\phi)$, and $\braket{\Pi_x^2}_{\phi} = 1$, which implies HL sensitivity when inserted into Eq.~\eqref{eq:SME_EstimatorVariance}. It is worth noting that, in the absence of decoherence, a parity measurement also constitutes an optimal measurement for both CSSs and SSSs with respect to saturating the QCRB at $\phi=0$. However, in the case of a CSS, the expectation value of the parity operator is given by $\braket{\Pi_x}_{\phi} = \cos^N(\phi)$, which rapidly approaches zero as $\phi$ deviates from zero, but nonetheless achieves the SQL around $\phi=0$, which is also the case for a SSS. 

Figure~\ref{fig:GHZ_SME} illustrates two primary limitations of the GHZ state. First, the phase range over which the signal can be unambiguously estimated is reduced by a factor of $N$, a direct consequence of the $N$-fold enhancement in oscillation frequency. Second, as shown in Fig.~\ref{fig:CSS_MLE_SME}, a larger number of measurement repetitions is required to saturate the CRB when using GHZ states compared to CSS. This can be understood by recognizing that a GHZ state effectively behaves as a single “super-atom” with an enhanced phase sensitivity scaling as $N$. Consequently, the effective number of independent samples is simply the number of measurements $r$. In contrast, for a CSS, each of the $N$ atoms contributes independently, resulting in an effective sample size of $N r$, and thus requiring fewer repetitions to reach the CRB. Again, the squared estimation error dips below the SQL near $\phi = \pm \pi/N$, which can be attributed to the fact that at these points the GHZ state becomes an eigenstate of the measurement operator, eliminating measurement noise.

The previous discussion highlights a limitation of the frequentist approach: a measurement that is optimal in the sense of saturating the CRB may nevertheless be inferior to another measurement that achieves the same bound. An example is provided by a CSS with a parity measurement. In this case, the encoded phase must be close to $\phi = 0$ in order to saturate the QCRB, whereas a measurement of $J_y$ achieves saturation over a much broader range of phase values. This naturally raises the question of how to identify measurements that maintain high sensitivity across a wide interval of phase values.

\section{Bayesian quantum phase estimation}
\label{sec:Bayesian}
For many practical applications (see Sect.~\ref{sec:frequencyEstimation}) it is more important to achieve accurate phase estimates with a limited number of measurement repetitions and to identify measurement protocols that yield low estimator variance across a broad range of phase values than it is to achieve the HL. One promising approach to address these requirements is the Bayesian framework for phase estimation.

In Bayesian quantum phase estimation ~\cite{buvzek1999optimal, macieszczak2014bayesian, jarzyna2015true, li2018frequentist, rubio2019quantum, kaubruegger2021quantum, kielinski2025bayesian}, the central premise is that the experimenter possesses initial knowledge about the parameter to be estimated which is captured in the prior probability density $p(\phi)$. This assumption is, in fact, not fundamentally different from frequentist approaches, where successful estimation relies on some prior knowledge to identify the correct interval of unambiguous estimation.

Upon obtaining a measurement outcome $\mu$, the uncertainty about the phase is updated according to Bayes’ theorem, yielding the posterior distribution
\begin{equation}
p(\phi|\mu) = \frac{p(\mu|\phi)p(\phi)}{p(\mu)},
\end{equation}
where $p(\mu|\phi)$ is the conditional probability of observing outcome $\mu$ given phase $\phi$, and 
\begin{equation}
    p(\mu) = \int_{-\infty}^{+\infty} d\phi\ p(\mu|\phi)\,p(\phi)
    \label{eq:evidence}
\end{equation}
is the probability to observe the measurement outcome $\mu$ given the prior knowledge about $\phi$. This posterior encapsulates the uncertainty about the phase after the measurement.

To quantify the average uncertainty after a single measurement, the figure of merit in the Bayesian framework is the mean squared error (MSE), defined in Eq.~\eqref{eq:MSE}, and averaged over the prior distribution as $\int_{-\infty}^{+\infty} d\phi\, \varepsilon(\phi)\, p(\phi)$. 

As a first step, the average MSE can be minimized with respect to the estimator. This minimum is achieved by the minimum mean squared error (MMSE) estimator, which corresponds to the  mean of the posterior distribution:
\begin{equation}
\varphi^*(\mu) = \int_{-\infty}^{+\infty} d\phi\, \phi\, p(\phi|\mu).
\label{eq:post_mean}
\end{equation}
Substituting this optimal estimator into the average MSE yields:
\begin{eqnarray}
\Delta_{\rm post}^2 &=& \int_{-\infty}^{+\infty} d\phi \sum_{\mu} \left(\phi- \varphi^*(\mu) \right)^2 p(\mu|\phi)\, p(\phi) \notag \\
&=& \sum_{\mu} p(\mu) \int_{-\infty}^{+\infty} d\phi \left(\phi- \varphi^*(\mu) \right)^2 p(\phi|\mu)\notag \\
&=&\sum_{\mu} p(\mu) \int_{-\infty}^{+\infty} d\phi \left(\phi- \int_{-\infty}^{+\infty} d\phi'\, \phi'\, p(\phi'|\mu) \right)^2 p(\phi|\mu)
\label{eq:PostVar}
\end{eqnarray}
The last expression shows that this quantity corresponds to the variance of the posterior distributions, averaged over all possible measurement outcomes. We therefore refer to $\Delta^2$ as the posterior variance.

\subsection{Optimal quantum interferometer}

The objective of Bayesian quantum phase estimation is to identify entangled probe states and measurement strategies that minimize the posterior variance for a given prior distribution. In contrast to the frequentist approach, establishing fundamental sensitivity limits in the Bayesian framework for arbitrary priors is generally nontrivial. However, an efficient numerical method to perform this minimization was introduced in Ref.~\cite{macieszczak2014bayesian}, based on two key observations.

First, for a fixed quantum state $\rho$, the posterior variance is minimized by measuring an operator $\mathcal{M}^*$ that satisfies the anticommutator equation
\begin{equation}
    \rho'=\frac{\M^*\overline{\rho} + \overline{\rho}\M^*}{2}, 
\end{equation}
where $\overline{\rho} = \int_{-\infty}^{+\infty} d\phi\,  U_{\phi}\,\rho\, U^{\dagger}_{\phi}\, p(\phi)$ and 
$\rho'=\int_{-\infty}^{+\infty} d\phi\, \phi\, U_{\phi}\,\rho\, U^{\dagger}_{\phi}\, p(\phi)$. 
This expression bears a formal resemblance to the SLD equation [Eq.~\eqref{eq:SLD}], and likewise, the optimal measurement operator can be written as
\begin{equation}
    \M^*=2\sum_{k, l}\frac{\braket{\overline{\psi_{k}}|\rho'|\overline{\psi_{l}}}}{\overline{p_k}+\overline{p_l}}\ket{\overline{\psi_{ k}}}\bra{\overline{\psi_{ l}}},
    \label{eq:OptimalMeasurement}
\end{equation}
where $\overline{\rho}= \sum_{k}\overline{p_k}\ket{\overline{\psi_k}}\bra{\overline{\psi_k}}$ is the spectral decomposition of the averaged density matrix. This procedure simultaneously optimizes over both the estimator and the measurement basis. Such that the minimal posterior variance for a given initial density matrix is given by 
\begin{equation}
    \Delta_{\rm post}^2\geq \overline{\phi^2}-\left(\overline{\phi^1}\right)^2 - \tr\left[\overline{\rho} \left(\M^*\right)^2\right],
    \label{eq:QuantumPostVar}
\end{equation}
with $\overline{\phi^k}=\int_{-\infty}^{+\infty}d\phi\, \phi^k p(\phi)$.

Second, for a fixed measurement operator $\M$, the optimal probe state $\ket{\psi^*}$ that minimizes the posterior variance can be obtained by solving the eigenvalue problem:
\begin{equation}
    \ket{\psi^*} = \arg\min_{\ket{\psi}} \int_{-\infty}^{+\infty}d\phi\, \bra{\psi}\M_{\phi}^2-2\phi \M_{\phi} \ket{\psi},
    \label{eq:optimal_state}
\end{equation}
where $\M_{\phi}=U_{\phi}\M U^{\dagger}_{\phi}$.

Remarkably, an alternating optimization over the probe state and the measurement operator converges rapidly to an optimal solution~\cite{macieszczak2013quantum,macieszczak2014bayesian}. This approach enables the determination of the fundamental sensitivity limits of the Bayesian cost function for a given prior. In the following, we refer to this ultimate quantum-enhanced sensing strategy as the optimal quantum interferometer (OQI).

\subsection{Phase operator}
Under some assumptions, it is possible to obtain an analytical solution for an optimal Bayesian sensor for a prior $p(\phi) = \frac{1}{2\pi}\Theta(\pi - |\phi|)$, where $\Theta$ is the Heaviside step function. However, to make this problem analytically tractable the MSE has to be replaced by a $\pi$-periodic function such as $4\sin^2\left(\frac{\phi-\varphi(\mu)}{2}\right)$. This is different from the squared error, but for $|\phi-\varphi(\mu)|\ll 1$ we have $4\sin^2\left(\frac{\phi-\varphi(\mu)}{2}\right)\simeq (\phi- \varphi)^2 $. 

In this scenario the optimal measurement is a measurement of the phase operator ~\cite{holevo2011probabilistic, pegg1988unitary, sanders1995optimal, derka1998universal}, which is defined as  
\begin{equation}
    \Phi=\sum_{k=-N/2}^{N/2} \Phi_k\ket{\Phi_k}\bra{\Phi_k}. 
\end{equation}
Herre $\Phi_k=\frac{2\pi k}{N+1}$ and the phase states $\ket{\Phi_k}=e^{-i\phi_kJ_z}\ket{\Phi_0}$ which form an orthogonal basis, and are obtained by rotating the seed phase state $\ket{\Phi_0}=\frac{1}{\sqrt{N+1}}\sum_{M=-N/2}^{N/2}\ket{M}$ by equidistant amounts around the $z$-axis. In analogy to watch each of the $N+1$ phase states can be interpreted as a distinct pointer state. As the initial sensor state—initially overlapping predominantly with $\ket{\Phi_0}$—undergoes a rotation by an angle $\phi$ about the $z$-axis, it will acquire the largest overlap with the phase state $\ket{\Phi_k}$ whose associated phase $\Phi_k$ is closest to the encoded phase $\phi$. It is therefore not surprising that the posterior variance of the phase operator, minimized over all possible initial states, is given by $\pi^2/(N+1)^2$. This bound is asymptotically saturated for large $N$ by the so-called sine state ~\cite{buvzek1999optimal, berry2000optimal}, defined as
\begin{equation}
    \ket{\psi_{\rm sin}}=\sqrt{\frac{2}{N+1}}\sum_{M}\sin\left(\frac{\pi(M+1/2)}{N+1}\right)\ket{M}. 
\end{equation}
This level of sensitivity coincides with the asymptotic bound for Bayesian phase estimation in the presence of priors supported on the interval $[-\pi, \pi]$. In this regime, the ultimate precision limit is given by the so-called $\pi$-corrected HL, which reads ~\cite{berry2000optimal, jarzyna2015true, gorecki2020pi}
\begin{equation}
\Delta^2_{\pi\rm HL}=\frac{\pi^2}{N^2}.
\end{equation}
\subsection{Bayesian Cramér-Rao bound}
\label{sec:BCRB}
To gain a more intuitive understanding of the fundamental limits imposed on the posterior variance, it is instructive to consider the Bayesian CRB ~\cite{gill1995applications, li2018frequentist, jarzyna2015true} for a quantum sensor prepared in an initial state $\rho$ and measured in a basis $\ket{\mu}$. The bound is given by
\begin{equation}
    \Delta_{\rm post}^2 \geq \frac{1}{\overline{F_{\rho,\ket{\mu}}} + \mathcal{I}}, 
\end{equation}
where 
\begin{equation}
\overline{F_{\rho,\ket{\mu}}}=\int_{-\infty}^{+\infty}d\phi \, F_{\rho_{\phi},\ket{\mu}}\,p(\phi)    
\end{equation}
is the classical FI [Eq.~\eqref{eq:FisherInformation}] averaged over the prior and 
\begin{equation}
    \mathcal{I}=\int_{-\infty}^{+\infty}d\phi \frac{1}{p(\phi)}\left(\frac{\partial}{\partial \phi}p(\phi)\right)^2, 
\end{equation}
is the FI of the prior density, i.e. quantifies how much knowledge we possess about the prior before the measurement. This expression naturally extends to the quantum Bayesian CRB if the classical FI in the average is replaced by the QFI. 

By construction, the average posterior variance in Eq.~\eqref{eq:PostVar} contains contributions from both the prior knowledge and the measurement outcomes. To isolate the contribution due solely to the measurement—i.e., the average estimator variance—the Bayesian CRB suggest that one can subtract the prior information from the posterior precision. This leads to the definition of an average estimator variance $\Delta$ ~\cite{leroux2017line, kaubruegger2021quantum}, given by
\begin{equation}
    \frac{1}{\Delta^2}=\frac{1}{\Delta_{\rm post}^2}-\mathcal{I}.
    \label{eq:AverageEstVar}
\end{equation}
Since the average posterior variance satisfies the bound $\Delta^2\geq1/\overline{F}$, the limits from frequentist phase estimation—namely, the SQL and the HL also apply to this average estimator variance.

\begin{figure}[b] 
   \centering
   \includegraphics[]{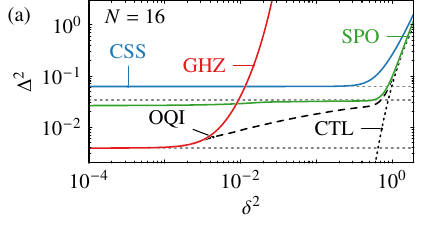}\\
   \includegraphics[]{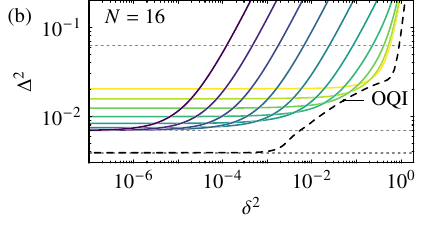}
   \caption{Average estimator variance $\Delta^2$ as a function of the prior variance for a $N = 16$ sensor. (a) Comparison of a coherent spin state (CSS) measured in the $y$-basis, a GHZ-state measured in the basis of the parity operator $\Pi_x$, a sine state measured in the basis of the phase operator (SPO), the optimal quantum interferometer (OQI) and the coherence time limit (CTL). The gray dotted horizontal lines correspond to $1 / N$, $\pi^2/N^2$, and $1 / N ^2$. (b) Comparison of  spin-squeezed states measured in the $y$-basis. The states are obtained as the ground states of the spin-squeezing Hamiltonian [Eq.~\eqref{eq:SqueezingParent}] for various values of the parameter $\omega/\chi$. The gray dotted horizontal lines correspond to $1 / N$, $2/(N^2+2N)$, and $1 / N ^2$.}
   \label{fig:DynamicRange}
\end{figure}

In many practical scenarios, prior knowledge about the phase can be well approximated by a normal distribution,
\begin{equation}
    p_{\delta}(\phi)=\frac{1}{\sqrt{2\pi\delta^2}}e^{-\frac{\phi^2}{2\delta^2}}, 
\end{equation}
e.g. as a result of performing a number of preliminary measurements—yielding a Gaussian prior via the central limit theorem—or when tracking a time-varying signal modeled as a Gaussian random process. The latter is particularly relevant for atomic clocks, as discussed in Sect.~\ref{sec:frequencyEstimation}.
The FI of a Gaussian prior is $\mathcal{I}=\frac{1}{\delta^2}$. If the prior distribution has a nonzero mean, this offset can be compensated by applying an additional rotation around the $z$-axis, either before or after the phase encoding, by an angle equal to the negative of the mean. Therefore, without loss of generality, we restrict the following discussion to prior distributions with zero mean.

For a Gaussian prior distribution, the posterior variance can be further related to the classical FI by considering the limit $\delta \rightarrow 0$. In this regime, the posterior variance admits an expansion ~\cite{macieszczak2013quantum, vasilyev2024optimal}
\begin{equation}
\Delta^2_{\rm post} = \delta^2 - \delta^4 F_{\rho_0} + \mathcal{O}(\delta^6),
\end{equation}
in terms of the classical FI associated with the initial state and the chosen measurement. Consequently, the average estimator variance in this regime is given by $\Delta^2 = 1 / F_{\rho_0}$. This also implies that, minimizing the posterior variance in the limit $\delta \rightarrow 0$ is equivalent to maximizing the classical FI of the sensor. 

Personick ~\cite{personick1971application} showed that, for single-parameter unitary estimation, the posterior variance can be minimized by a von Neumann measurement. This result implies that there always exists a projective measurement that saturates the QCRB. However, this does not preclude the possibility that a POVM with comparable performance may be more practical to implement in a given experimental setting.

Furthermore, this equivalence allows Eq.~\eqref{eq:optimal_state} to be used to identify the quantum state that maximizes the classical FI for a given measurement. Notably, this task constitutes a nontrivial optimization problem when addressed purely within the frequentist framework.

\subsection{Dynamic range}

The comparison of different sensing strategies in Fig.~\ref{fig:DynamicRange} (a) illustrates several key features previously discussed. Notably, a GHZ-state-based interferometer is capable of achieving the HL, but only when the prior distribution has a variance $\delta^2 \lesssim 1/N^2$. In contrast, an unentangled sensor employing a CSS and single-atom $J_y$ measurements achieves the SQL for $\delta^2 \lesssim 1$. However, for broader priors, the best sensitivity is attained by a sensor utilizing a sine state and a phase operator measurement. This highlights the importance of optimizing not only the quantum state but also the measurement basis in order to extend the range of phase values over which precise estimation is possible.

Another observation is that the average estimator variance of all considered sensors increases rapidly once $4\delta^2 \gtrsim 1$. This behavior can be attributed to the fact that, for sensors composed of a single ensemble of identically prepared atoms, measurement outcomes can only uniquely resolve phase values within the interval $[-\pi, \pi]$. When phase slips occur beyond this range, the estimator variance rises sharply. The probability of phase slips is  $1 - \rm{erf}(\pi/\sqrt{2}\delta)$. Consequently, the increase in average estimator variance for both the OQI and the sine-state sensor can be well approximated by the expression
$4\pi^2\left(1 - \rm{erf}\left(\frac{\pi}{\sqrt{2}\delta}\right)\right)$,
as indicated by the black dashed line.

In the following, we refer to the range of phase values—characterized by the prior variance $\delta^2$—over which a sensor can maintain a given average estimator variance as the dynamic range of the sensor at that specific sensitivity.

Equipped with the tools to assess the dynamic range of a quantum sensor, we now revisit the SSSs analyzed in Fig.~\ref{fig:SSS}, with the corresponding evaluation presented in Fig.~\ref{fig:DynamicRange} (a). These results reinforce the earlier intuition that increasing the degree of squeezing comes at the cost of a reduced range over which the best sensitivity can be maintained. This trade-off is especially pronounced for the maximally SSS, which has a dynamic range significantly smaller than that of the GHZ state, rendering it practically ineffective as a sensor.

To exploit the metrological potential of such a highly squeezed state, it is necessary to adapt the measurement strategy. Indeed, when the phase is measured in the eigenbasis of the phase operator, the dynamic range of the maximally SSS can be extended, reaching a level comparable to that of the GHZ state.

This observation highlights a limitation of the frequentist approach ~\cite{vasilyev2024optimal}. Both the $J_y$ measurement and the phase operator measurement saturate the QCRB for the maximally SSS. However, the dynamic ranges resulting from these two measurement strategies differ significantly. Only through the calculation of the posterior variance within the Bayesian framework can these practically relevant distinctions be captured, underscoring the importance of considering measurement-dependent performance criteria beyond asymptotic bounds.

\subsection{Extending dynamic range}
\label{sec:extendingDynamicRange}
Having established the dynamic range as an important performance metric of a quantum sensor, a natural question arises: how can the dynamic range be extended? This question is closely linked to the choice of measurement strategy. In the following, we provide an overview of several approaches that have been proposed to enhance the dynamic range, and evaluate their performance in terms of the resulting average estimator variance.

\subsubsection{Dual quadrature measurement}

A straightforward approach to increasing the dynamic range of a quantum sensor is to divide the total ensemble of atoms into 2 sub-ensembles and perform a measurement in the basis of $\frac{J_x+J_y}{\sqrt{2}}$ in one of the ensembles and in the basis of $\frac{J_x-J_y}{\sqrt{2}}$ in the second ensemble ~\cite{rosenband2013exponential, borregaard2013efficient, li2022improved}. Combining the two measurement outcomes the phase can be unambiguously tracked over the interval $[-\pi,+\pi]$. This is referred to as a dual quadrature measurement and can be extended to more ensembles that are measured in different bases ~\cite{shaw2024multi, zheng2024reducing, hainzer2024correlation}. 

\subsubsection{Attenuated phases}
\label{sec:attenuated_phase}
A strategy to further extend the dynamic range is to divide the atoms into $n$ ensembles, such that the atoms in the $k$-th ensemble acquires an attenuated phase of $\phi/2^{k-1}$. This scaling of the encoded phase can be implemented by allowing the atoms in each ensemble to interact with the signal that generates the phase for a proportionally reduced interaction time. Each of these sub-ensembles is then split in half, and a dual quadrature measurement is performed. In the limit of small prior width, the phase estimation variance for each ensemble prepared in a CSS is given by $\Delta_k^2 = \frac{n}{N} 2^{2(k-1)}$, since each ensemble contains $N/n$ atoms and the encoded phase is attenuated by a factor $1/2^{k-1}$, which increases the variance quadratically. The resulting bound on the total average estimator variance is therefore
$\Delta^2 \leq \left( \sum_{k=1}^n \frac{1}{\Delta_k^2} \right)^{-1} = \frac{3n}{4(1 - 4^{-n})}  \frac{1}{N}$,
which exceeds the SQL.

\begin{figure}[b] 
   \centering
   \includegraphics[]{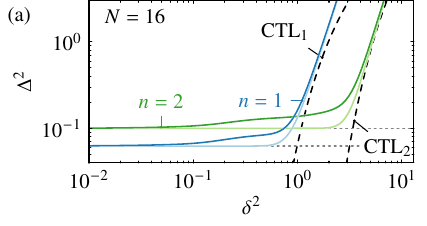}\\
   \includegraphics[]{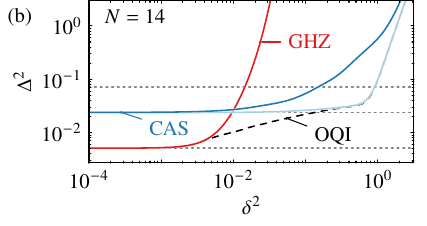}
   \caption{Comparison of the average estimator variance $\Delta^2$ as a function of the prior variance $\delta^2$ for quantum sensors composed of $N$ atoms. (a) The ensemble is divided into $n$ sub-ensembles of $N/n$ atoms, with each sub-ensemble interrogated by a phase that is attenuated by a factor of 2 relative to the previous one. Dark-colored curves indicate the performance of dual-quadrature measurements applied to each sub-ensemble, while light-colored curves represent the minimum average estimator variance obtained by optimizing over all possible measurement strategies. The horizontal gray dotted lines mark the values $1/N$ and $8/(5N)$, respectively, and the vertical black dashed lines denote the coherence time limit (CTL), defined as ${\rm CTL}n = (2n\pi)^2\left(1 - \int_{-n\pi}^{+n\pi} d\phi\, p_{\delta}(\phi)\right).$ (b) The atoms are divided into six sub-ensembles, consisting of two ensembles each with $1$, $2$, and $4$ atoms. Each sub-ensemble is initialized in a GHZ state, and a dual-quadrature parity measurement is applied to pairs of ensembles with identical atom numbers (dark blue). The light blue curve represents the average estimator variance minimized over all possible measurements. For reference, the performance of an $N$-atom GHZ sensor and the optimal quantum interferometer are also shown. The horizontal gray dotted lines mark the values $1/N$ and $6/(N^2+4N)$, and $1/N^2$. }
   \label{fig:DR_Multi}
\end{figure}

As shown in Fig.\ref{fig:DR_Multi}(a), the average estimator variance reaches its minimum in the regime of narrow prior distributions. At the same time, the dynamic range can be extended and approaches the coherence time limit associated with unambiguous phase estimation within the interval $[-n\,\pi, +n\,\pi]$. This extension comes at the cost of an increased average estimator variance relative to the small prior limit. These results further underscore the potential of optimizing the measurement basis to minimize the average estimator variance across the extended dynamic range. In particular, the variance obtained by optimizing over all possible measurements falls below that achievable with a dual-quadrature measurement scheme at the larger prior widths. While the theoretically achievable sensitivity can be computed numerically, the explicit form of the measurement strategy required to realize this performance is not immediately apparent; see Sect.~\ref{sec:approx_opt_meas} for a discussion on how to approximate the optimal measurement basis.

\subsubsection{Amplified phases}
\label{sec:GHZ_Cascade}

\begin{figure*}[t] 
   \centering
   \includegraphics[width=\textwidth]
   {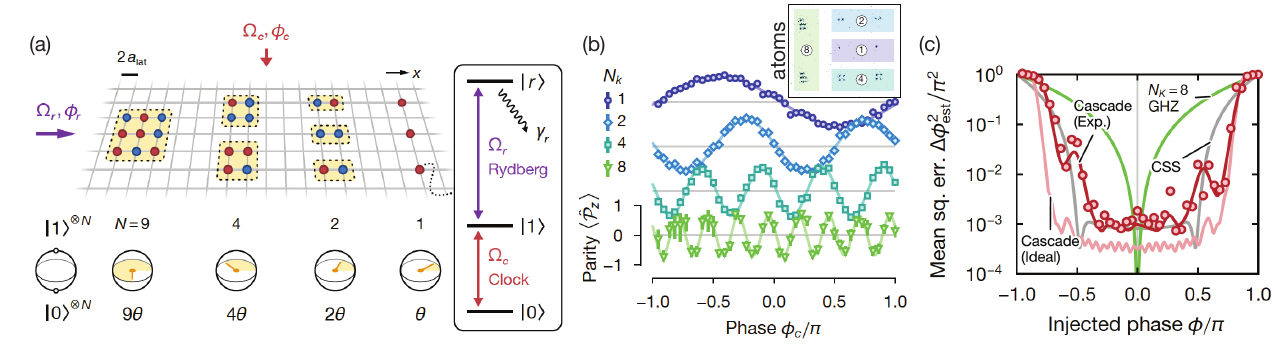}
   \caption{(a) Preparation of an array of atomic ensembles in GHZ states of varying sizes (‘cascaded GHZ states’), where yellow regions denote atomic clusters forming individual GHZ states. Each GHZ state size accumulates phase at a rate proportional to its particle number. Inset: GHZ states are generated on the optical clock transition of strontium, with entanglement established via excitation to Rydberg states. (b) Measured parity signals for GHZ states of different sizes $N_k$, showing oscillations whose frequency increases with ensemble size. Inset: atom array image of the sample employed to generate the cascaded GHZ states. (c) Performance of the mean squared error of the cascaded GHZ states compared with alternative input states. Adapted from Ref.~\cite{cao2024multi}.}
   \label{fig:cao_fig}
\end{figure*}

An alternative strategy for extending the dynamic range of quantum sensors involves employing ensembles that accumulate phase at an amplified rate ~\cite{berry2009perform,higgins2009demonstrating,kaftal2014usefulness,kessler2014heisenberg,direkci2024heisenberg}. As discussed in Sect.~\ref{sec:GHZ_state}, this can be realized using GHZ states with increasing atom number. Here, we illustrate this concept through a cascade configuration consisting of ensembles grouped in pairs: each pair contains two ensembles of equal atom number, with subsequent pairs doubling the atom number relative to the previous pair. This design supports dual quadrature parity measurements of the form $\Pi_\pm = \prod_{k=1}^N \left(\sigma_x^{(k)} \pm \sigma_y^{(k)}\right)/\sqrt{2}$, enabling unambiguous phase estimation over a range $[-\pi, +\pi]$.

The average estimator variance for this configuration is lower bounded by the QFI of the cascaded GHZ-state
$\Delta^2 \geq \left( 2\sum_{k=1}^n 4^{k+1} \right)^{-1} = \frac{6}{N^2 + 2N}$.
As shown in Fig.\ref{fig:DR_Multi} (b), the GHZ-state cascade saturates the lower bound in the regime of narrow prior distributions. However, when using the dual quadrature measurement protocol described in Ref.~\cite{kessler2014heisenberg}, the full dynamic range cannot be exploited without a degradation in precision. Remarkably, when combined with the optimal measurement strategy, the GHZ-cascade achieves performance on par with the OQI for prior variances $\delta^2 \lesssim 1$, allowing for both substantial dynamic range and entanglement-enhanced sensitivity. A detailed analysis of GHZ-cascade protocols—including the optimization of ensemble sizes, repetition numbers, and adaptive measurement schemes—can be found in Ref.~\cite{direkci2024heisenberg}.

Recent experimental realizations of cascaded GHZ states in tweezer-based optical atomic clocks have established this approach as a promising direction for quantum-enhanced metrology ~\cite{cao2024multi, finkelstein2024universal}. In particular, Cao et al. ~\cite{cao2024multi} demonstrated a sub-SQL measurement in an atom-laser comparison with a short dark time of 3 ms using 9 copies of 4-particle GHZ states, constrained primarily by residual atom-laser frequency fluctuations. In a proof-of-principle experiment, a cascaded GHZ protocol employing GHZ states with up to 8 atoms—formed from a total of 30 atoms distributed among different entangled state sizes—enabled an extended dynamic range (see Fig.~\ref{fig:cao_fig}). However, preparation errors led to reduced contrast in the interferometer, preventing performance below the SQL.

\subsubsection{Weak measurements}
Another strategy to extend the dynamic range is to perform weak measurements sequentially during the phase-acquisition process. Such measurements do not extract the full information that a conventional projective measurement would provide; instead, they only weakly perturb the quantum state without collapsing its wavefunction. The strength of this perturbation depends on the amount of information extracted, allowing one to track the accumulated phase and suppress phase-slip errors ~\cite{gefen2018quantum, cujia2019tracking, pfender2019high, cohen2020achieving, tratzmiller2020limited}. This approach has been studied in cavity-based atomic clocks ~\cite{shiga2012locking, borregaard2013near}, where the weak measurement is implemented by coupling the atoms to a bosonic cavity mode and subsequently measuring the cavity output. More recently, a related concept has been explored in systems where weak measurements are realized by entangling the clock atoms with ancillary atoms, which are then measured ~\cite{direkci2025extending}. 

\subsection{Approximating optimal measurements}
\label{sec:approx_opt_meas}

\begin{figure*}[t] 
   \centering
   \includegraphics[width=\textwidth]
   {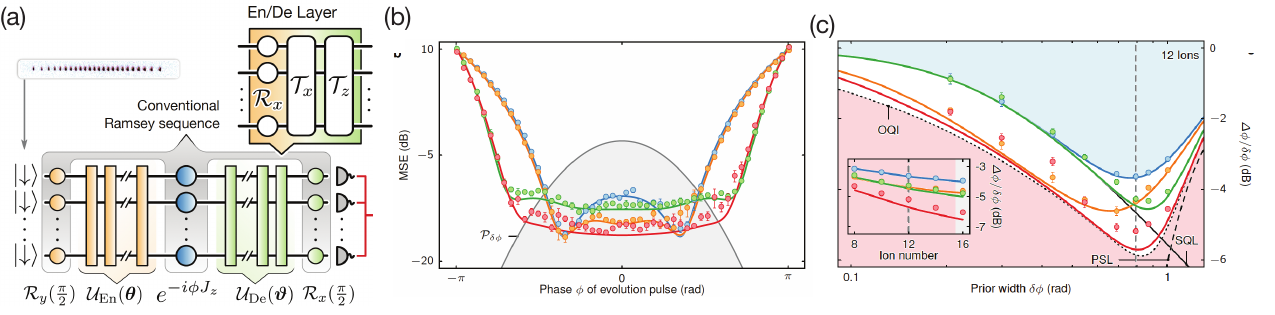}
   \caption{(a) A programmable quantum sensor implementing a generalized Ramsey sequence with entangling $U_{\rm en}(\bm{\theta})$ and decoding $U_{\rm de}(\bm{\vartheta})$ unitaries acting on N particles. Both unitaries are constructed from repeating sequences of sensor resource gates $\mathcal{R}$ and $\mathcal{T}$, parameterized by the parameter vectors $\{\bm{\theta},\bm{\vartheta}\}$. (b) Mean squared error for $N=16$ quantum sensors optimized under a Gaussian prior distribution of width $\approx 0.79$ (gray background). Colored lines denote: an unentangled sensor without entangler and decoder (blue), a spin-squeezed sensor without decoder (orange), a sensor with entangler but no decoder (green), and a fully optimized sensor with both entangler and decoder (red). (c) Ratio of posterior to prior standard deviation as a function of the prior width for the same sensors as in (b). The blue shaded region denotes the performance achievable without entanglement, while the red shaded region is inaccessible to single-shot measurement schemes; its boundary corresponds to the OQI limit. The OQI, the SQL, and the phase slip limit are indicated by black lines. Inset: scaling of the ratio with particle number at the prior width marked by the dashed line. Adapted from Ref.~\cite{marciniak2022optimal}.}
   \label{fig:ions_variational}
\end{figure*}

As demonstrated in several preceding examples, optimizing the measurement basis of a quantum sensor can significantly enhance its dynamic range. While the optimal measurement can, numerically be determined according to Eq.~\eqref{eq:OptimalMeasurement}, this does not necessarily imply that it is straightforward to implement in an experimental setting. In theory, any projective measurement in a desired basis can be realized by applying an appropriate unitary decoder $U_{\rm de}$ prior to a projective measurement in the $z$-basis. However, constructing such a decoder in a systematic and experimentally feasible way remains a non-trivial challenge for generic target measurements.

One practical approach to this problem is to parameterize the decoding unitary using a class of variational quantum circuits ~\cite{cerezo2021variational}, as illustrated in Fig.~\ref{fig:ions_variational}. Specifically, the decoder can be expressed as
\begin{equation}
U_{\rm de}(\bm{\theta}) = e^{-i H_R^{(d)} \theta_d} \cdots e^{-i H_R^{(2)} \theta_2} e^{-i H_R^{(1)} \theta_1},
\end{equation}
where ${H_R^{(i)}}$ denotes a set of native resource Hamiltonians accessible with the given quantum sensor, and $\boldsymbol{\theta} = (\theta_1, \ldots, \theta_d)$ are tunable variational parameters. The depth $d$ defines the number of sequential layers in the circuit. These parameters are optimized to approximate the desired decoding unitary as accurately as possible.

Rather than attempting to match a specific target unitary, it is often more effective to directly minimize a metrologically relevant cost function ~\cite{shen2013efficient, kaubruegger2019variational, koczor2020variational}, such as the average estimator variance. This approach has been used to identify optimal decoders tailored to a given prior distribution~\cite{kaubruegger2021quantum}, while simultaneously optimizing a variational entangler to prepare a suitable entangled initial state. A notable experimental demonstration of this strategy has been realized on a trapped-ion quantum processor with up to 26 qubits ~\cite{marciniak2022optimal}.

Given current hardware limitations, many quantum sensing platforms cannot implement deep variational circuits, prompting the question of what types of decoding operations can be effectively realized with shallow circuits ~\cite{schulte2020ramsey, thurtell2024optimizing, scharnagl2023optimal, liu2025enhancing}.

This discussion also connects to a broader class of protocols in which the decoding operation is employed not solely to increase the dynamic range, but also to reduce the sensor’s susceptibility to detection noise ~\cite{davis2016approaching, frowis2016detecting, nolan2017optimal, haine2018using, colombo2022time, liu2022nonlinear, li2023generalized, liu2023cyclic}, or to enable the sensor to saturate the QCRB ~\cite{macri2016loschmidt, volkoff2022asymptotic}.

\subsubsection{Adaptive measurement protocols}
An alternative direction is to explore adaptive measurement protocols ~\cite{chabuda2016quantum, demkowicz2017adaptive, pang2017optimal, pirandola2017ultimate, pezze2020heisenberg, wan2022bounds,direkci2024heisenberg}, in which portions of the system are measured sequentially, and the outcomes of earlier measurements are used to update the measurement basis for subsequent measurements. This approach is particularly suitable in scenarios such as those discussed in Sect.~\ref{sec:extendingDynamicRange}, where the system is partitioned into multiple sub-ensembles.

Although adaptive protocols do not provide an asymptotic enhancement in precision in the frequentist regime \cite{kurdzialek2023using}, they can still be highly effective in practice. In particular, they offer a pragmatic route to implement near-optimal measurements without requiring complex operations on the full quantum system.

In Ref.~\cite{direkci2024heisenberg}, a protocol is considered in which the sensing atoms are divided into multiple sub-ensembles that interact with the signal for an identical duration but are measured sequentially. Each measurement can be adapted based on the outcomes of the preceding ones, enabling sensitivities that approach those of optimal collective measurements for a cascade of GHZ-states. 

A related strategy considers adaptive protocols where the atoms sequentially interact with the signal, allowing for dynamic modification of the initial state or real-time adaptation to time-dependent signals. However, determining precision bounds in such settings is computationally challenging: the numerical complexity of computing the fundamental sensitivity limit scales exponentially with the number of interrogation steps, restricting tractable analysis to systems with only a few atoms and time steps ~\cite{chabuda2016quantum}.

\section{Quantum Frequency estimation}
\label{sec:frequencyEstimation}
In the previous section, we introduced the principles of quantum phase estimation, discussed its fundamental limits, and motivated the concept of dynamic range. This emphasis arises from the fact that, in most practical applications of quantum sensors, the primary quantity of interest is not the accumulated phase
$\phi = \int_0^T dt\, \omega(t)$,
averaged over the interrogation time $T$ during which the sensor atoms interact with the signal, but rather the frequency $\omega(t)$ that generates the phase.

Typically, it is assumed that the frequency remains approximately constant during the interrogation period, such that $\phi/T$ provides a reliable estimate of the instantaneous frequency $\omega(T)$ at the end of the cycle.  Under this assumption, the variance of the frequency estimate is related to the phase estimation variance by
$\Delta^2_{\omega} = \frac{\Delta^2_{\phi}}{T^2}$,
making it advantageous to increase the interrogation time $T$ in order to improve frequency precision. However, longer interrogation times also increase the total accumulated phase uncertainty. Effective clock operation therefore requires a balance: interrogation times must be long enough to suppress frequency variance, yet short enough to ensure that the phase remains within the regime of unambiguous estimation. A detailed analysis of Bayesian frequency estimation in this context can be found in Ref.~\cite{kielinski2025bayesian}. 

\subsection{Atomic clocks}
\label{sec:clocks_theory}
A paradigmatic example of frequency estimation is provided by atomic clocks, in which a local oscillator is stabilized to an atomic transition frequency $\omega_0$. In this context, the frequency to be estimated is the detuning $\omega(t) = \omega_{\rm LO}(t) - \omega_0$ between the local oscillator frequency $\omega_{\rm LO}(t)$ and the atomic reference. Each clock cycle concludes with feedback that adjusts the local oscillator based on this estimate, thereby stabilizing the frequency of the local oscillator to the atomic transition frequency~\cite{ludlow2015optical}.

The instability of such clocks is typically characterized by the Allan deviation ~\cite{riley2008handbook}, which quantifies the stability of the  fractional frequency $y(t)= \frac{\omega(t)}{\omega_0}$ over different averaging times $\tau$. In terms of the average phase estimator variance Eq.~\eqref{eq:AverageEstVar}, the Allan deviation can be expressed as
\begin{equation}
\label{eq:instability}
\sigma_y(\tau) = \frac{1}{\omega_0} \frac{\Delta_{\phi}}{T} \sqrt{\frac{T + T_{\rm D}}{\tau}},
\end{equation}
where $\omega_0$ is the atomic transition frequency, $T$ is the interrogation time, and $T_{\rm D}$ accounts for dead time during which the atoms are not interacting with the local oscillator, for instance, during state preparation, measurement, or feedback operations.

Monte Carlo oscillator simulations of the full clock feedback loop ~\cite{leroux2017line, schulte2020prospects,kaubruegger2021quantum, kielinski2025bayesian} demonstrate that the average phase estimation variance directly determines the resulting Allan deviation. This naturally raises the question of how to appropriately model the prior phase distribution used in calculating the average estimator variance. While one might initially ascribe the width of this prior, $\delta_\phi$, solely to the intrinsic fluctuations of the free-running local oscillator, it is, in practice, more strongly influenced by the residual noise of the stabilized local oscillator frequency. This residual noise depends sensitively on the specific interrogation, estimation, and feedback procedures employed. Once the feedback loop effectively locks the local oscillator to the atomic reference, the prior distribution of the phase approaches a stationary form. In this regime, the residual noise can be approximated as white noise and modeled by a Gaussian distribution with variance ~\cite{leroux2017line}
\begin{equation}
\delta_{\phi}^2 \propto \left(\frac{T}{T_{\rm C}}\right)^{2+\alpha}. 
\end{equation}
The underlying local oscillator noise manifests itself in the exponent $\alpha = -1, 0, 1$ which corresponds to white, flicker, and random-walk frequency noise, respectively, characterized by a power spectral density of the form $S(f) = h_\alpha f^{-1-\alpha}$. The quantity $T_{\rm C}$ denotes the characteristic coherence time of the local oscillator, defined implicitly by the condition $\sigma_{y,{\rm LO}}(T_{\rm C}) \, 2\pi \omega_0 (T+T_{\rm C}) = 1~{\rm rad}$, where $\sigma_{y,{\rm LO}}(T_{\rm C})$ is the Allan deviation of the uncorrected local oscillator, averaged over a single cycle of interrogation and dead time. For the purposes of this work, it is sufficient to note that the prior variance $\delta_\phi^2$ increases monotonically with the interrogation time $T$; a more detailed discussion can be found in Refs.~\cite{leroux2017line, kielinski2025bayesian}.

\subsection{Entanglement-Enhanced Sensing in Clocks}

We begin by assuming that the dead time between consecutive measurements is negligible compared to the interrogation time. Accounting for dead time— referred to as the Dick effect (see Sect.\ref{sec:key_challanges})—can alter the conclusions drawn here and restricts the operational regime in which atomic clocks benefit from entanglement ~\cite{schulte2020prospects}.

Ideally, it is advantageous to maximize the interrogation time $T$, up to the averaging time 
$\tau$ of interest. This is because each phase estimate is affected by quantum projection noise, and increasing $T$ reduces the number of such noisy estimates per averaging time.

When the interrogation time approaches the averaging time $T \rightarrow \tau$, the Allan deviation scales as $\sigma(\tau) \propto 1/\tau$. In contrast, in the limit $T \ll \tau$, the Allan deviation has a less favorable scaling of $\sigma(\tau) \propto 1/\sqrt{\tau}$. These asymptotic behaviors provide intuition for identifying the most suitable quantum sensing strategies under varying experimental constraints and averaging regimes.

If the objective is to achieve enhanced short-term stability at averaging times satisfying $(\tau/T_{\rm C})^{2+\alpha} \lesssim \pi^2/N^2$, a GHZ-state-based interferometer is optimal. In this regime, the interrogation time of a GHZ-interferometer can match the averaging time, and the phase can be estimated unambiguously with minimal noise.

For averaging times $\tau \lesssim T_{\rm C}$, approaching the coherence time of the local oscillator, optimal performance is obtained using sensors prepared in sine states and measured in the eigenbasis of the phase operator. Spin-squeezed states or optimized cascades of GHZ states can also offer improvements over clocks based on unentangled atoms in this regime.

Finally, in the regime $\tau > T_{\rm C}$, extending the interrogation time to match the averaging time requires alternative schemes such as the attenuated phase estimation protocol introduced in Sect.~\ref{sec:attenuated_phase}. Even in this regime, it remains advantageous to operate each sub-ensemble in a sine state and perform measurements in the eigenbasis of the phase operator, or alternatively to employ spin-squeezed initial states, in order to suppress projection noise within the sub-ensembles.

\subsection{Atomic coherence time and phase estimation}
\label{sec:atomicCoherenceTime}

\begin{figure}[b] 
   \centering
   \includegraphics[]{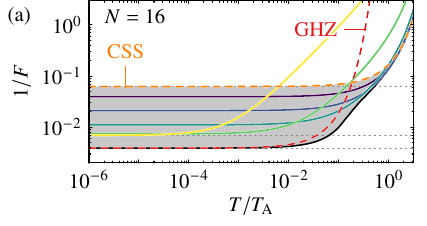}\\
   \includegraphics[]{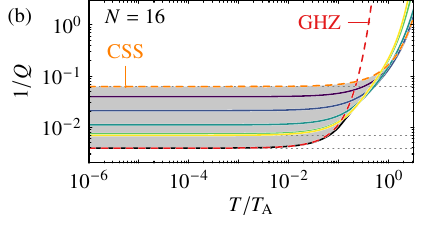}\\
   \caption{Comparison of the Cramér–Rao bound (a) and QCRB (b) on the phase estimator variance $\Delta_{\varphi}^2\geq 1/F\geq 1/Q$ after exposure to spontaneous emission for a duration $T$. For reference, the (Q)CRBs corresponding to a coherent spin state (CSS) and a GHZ state are also shown. The horizontal dotted lines correspond to $1/N,2/(N^2+2N), 1/N^2$. Colored curves correspond to spin-squeezed states— that are eigenstates of the Hamiltonian in Eq.\eqref{eq:SqueezingParent}—with increasing squeezing from purple to yellow. In the short-time limit $T/T_{\rm A}\!\to\!0$, the inverse (Q)FI reduces to the squared squeezing parameter. For all states, $F$ is evaluated with respect to projective measurements in the $y$-basis. }
   \label{fig:FI_SSS}
\end{figure}

If you have a nearly noiseless local oscillator, it may appear advantageous to continually increase the atomic interrogation time to improve precision. However, this strategy is ultimately constrained by the intrinsic coherence time of the atomic system. In what follows, we assume that all technical noise sources limiting coherent evolution are negligible and consider only the fundamental decoherence mechanism of spontaneous decay from the excited to the ground state of the reference transition, occurring at a rate $1/T_{\rm A}$, where $T_{\rm A}$ denotes the atomic coherence time and $\omega_0$ the transition frequency. In optical clocks, the rate of this decoherence mechanism can be conveniently determined through atom–atom comparisons ~\cite{marti2018imaging, young2020half}.

Increasing the interrogation time beyond the atomic coherence time yields diminishing returns, and no improvement in scaling with system size can be achieved ~\cite{huelga1997improvement}. A major challenge in this regime is that quantum states with large QFI are also typically more fragile and susceptible to decoherence. To make this trade-off explicit, we consider the evolution of the system under a Lindblad master equation of the form:

\begin{equation}
    \frac{\partial}{\partial t}\rho=\Lambda\circ\rho=\frac{1}{T_{\rm A}}\sum_{k=1}^N 2\sigma_-^{(k)}\rho\sigma_{+}^{(k)}-\sigma_{+}^{(k)}\sigma_{-}^{(k)}\rho - \rho\sigma_{+}^{(k)}\sigma_{-}^{(k)}.
\end{equation}

Since the phase-encoding operation commutes with the Lindblad operator describing spontaneous emission, the two processes can be applied sequentially at the theoretical level. Consequently, one may first let decoherence act on the initial state $\rho_0$ to obtain the state $\rho_T=e^{T\Lambda\circ}\rho_0$ and then evaluate the phase estimation FI according to Eq.~\eqref{eq:FisherInformation} and QFI according to Eq.~\eqref{eq:QFI} of the resulting mixed state as a measure of the sensitivity lost due to spontaneous emission.

Figure~\ref{fig:FI_SSS}(a) illustrates that SSSs with stronger initial squeezing show a more rapid degradation of FI under spontaneous emission, leading to a corresponding increase in the CRB for phase estimation. A surprising exception to this trend is the GHZ state, which retains its FI for a longer duration than strongly squeezed states. We emphasize that this behavior is specific to spontaneous emission and does not hold for other types of noise, and furthermore occurs only for intermediate system sizes. This robustness arises from the fact that, for a GHZ state, trajectories involving spontaneous emission events are distinguishable from those without errors. As a result, erroneous outcomes can be identified and excluded from the estimation process, thereby mitigating their detrimental impact on sensitivity ~\cite{kielinski2024ghz} (see Sect.~\ref{sec:robustness} for more detail). In this case, the improvement is possible because we evaluate the FI associated with a excitation number resolved measurement in the $y$-basis, instead of only considering two probabilities of either being in the $+$ or $-$ eigenspace of the $\Pi_x$ parity measurement discussed in Sect.~\ref{sec:GHZ_state}. 

A comparison with Fig.~\ref{fig:FI_SSS}(b) shows that decoherence effects can be mitigated by optimizing the measurement basis. This suggests a potentially promising route for enhancing the sensitivity of quantum metrology with SSSs at intermediate system sizes. Furthermore, we find that for nearly all interaction times, the ultimate sensitivity bound obtained following the method described in Ref.~\cite{macieszczak2013quantum} is saturated by either GHZ states, SSS, or the CSS. 

This discussion highlights a key point: even when employing multi-ensemble schemes to extend the dynamic range, the atomic coherence time imposes a fundamental limit. For example, in an attenuated phase estimation scheme (Sect.~\ref{sec:attenuated_phase}), where different sub-ensembles interact with the signal for varying durations, the sub-ensemble with the longest interrogation time should not exceed the atomic coherence time. Beyond this limit, it no longer contributes useful information and thus loses its utility. Conversely, for interrogation times approaching T $\lesssim T_{\rm A}$, entanglement no longer provides a scalable advantage in sensitivity. Nevertheless, for shorter interrogation times, such as those of the attenuated sub-ensembles that probe the signal for reduced durations, entanglement can still enhance precision. 

\subsection{Atomic coherence time and frequency estimation}
\label{sec:bandwidth}

\begin{figure}[b] 
   \centering
   \includegraphics[]{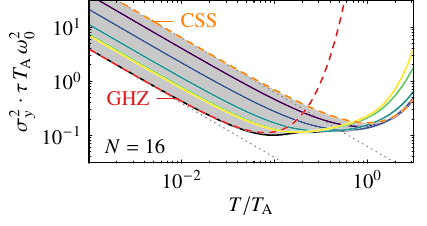}
   \caption{Quantum Cramér–Rao–bound (QCRB) Allan variance for frequency estimation in the presence of spontaneous emission. The Allan variance is obtained by rescaling the results of Fig.\ref{fig:FI_SSS}(b). The gray shaded region marks the parameter regime where the sensitivity achievable with entangled states exceeds that of a coherent spin state (CSS). The black line indicates the lower bound obtained by maximizing the quantum Fisher information in the presence of spontaneous emission over all possible quantum states. Dotted gray lines show the short-time asymptotic limits for a GHZ state $1/(N^2T)\cdot \tau\,T_{\rm A}$ and a CSS $1/(N\,T)\cdot \tau\,T_{\rm A}$. Colored curves correspond to spin-squeezed states—eigenstates of the Hamiltonian in Eq.\eqref{eq:SqueezingParent}—with increasing squeezing from purple to yellow.}
   \label{fig:QFI_SSS_f}
\end{figure}

Building on these observations, we now examine the advantages of employing entangled states for frequency estimation in the presence of a finite atomic coherence time $T_{\rm A}$. Although the task remains fundamentally one of phase estimation, a natural quantifier for frequency sensitivity is the Allan variance defined in Eq.\eqref{eq:instability}, which incorporates the time-dependent phase-estimator variance that increases with the interrogation time. For simplicity, we assume that the dead time $T_{\rm D}$ is negligible compared with the interrogation time $T$. Under this assumption, the Allan variance takes the form $\sigma^2_y=\frac{\Delta_{\phi}^2}{\omega_0^2\, T\, \tau}$. Figure~\ref{fig:QFI_SSS_f} shows the corresponding QCRB on the Allan variance as a function of the interrogation time. For interrogation times shorter than the atomic coherence time, entangled states can achieve a lower Allan variance than a CSS.

To achieve the smallest possible Allan variance, one should operate the sensor at the largest interrogation time for which the phase-estimation variance in Fig.~\ref{fig:FI_SSS} increases slower than linearly with $T$. A strictly linear increase offers no advantage, since the same sensitivity can be reached by choosing a smaller interrogation time and performing more repetitions within the averaging time $\tau$. In practice, this optimal operating point lies near the interrogation time $T\sim T_{\rm A} / N$ where the phase estimator uncertainty of a GHZ state begins to increase.

While our analysis focuses on a single system size, the maximum improvement achievable with entangled states over a CSS does not scale with $N$~\cite{huelga1997improvement, haase2018fundamental}. This finite advantage is further reduced when the dead time $T_{\rm D}$ is non-negligible, which is more severe at shorter interrogation times. Furthermore, $T_{\rm D}$ is often larger for entangled-state protocols due to the additional time required for state preparation and for implementing more complex measurement operations. As a result, the practical benefit of entanglement in frequency metrology lies in the increased measurement bandwidth: entangled states enable a given frequency precision to be reached at shorter interrogation times.

For any given quantum state, there exists an interrogation time that minimizes the Allan variance. As the degree of multipartite entanglement increases—from modestly SSSs to maximally entangled GHZ states—this optimal time shifts toward shorter values. This reflects a fundamental trade-off between sensitivity and bandwidth in entanglement-enhanced metrology. Entangled states can offer superior sensitivity at short interrogation times, but their heightened susceptibility to decoherence—most pronounced for strongly squeezed and GHZ states—limits their performance at longer times. Consequently, when bandwidth is not a consideration, entanglement does not yield a scalable improvement over the CSS.

\section{Current pursuits} 
\label{sec:current}

Up to this point, our discussion has focused primarily on theoretical considerations, in particular those arising from the fundamentally periodic nature of quantum evolution in finite systems—which constrains the dynamic range of a sensor—and from the unavoidable limitation imposed by spontaneous emission. While these effects set fundamental performance bounds, the practical realization of quantum sensors is subject to a broader range of experimental challenges. In practice, numerous additional factors can hinder performance improvements, including technical noise sources, finite control fidelities, imperfect state preparation and detection, and environmental perturbations specific to the sensing platform.

Addressing these challenges has been a central driver of experimental progress in the field. Over the past fifteen years, there have been ground-breaking demonstrations of entanglement-enabled precision gains~\cite{pezze2018quantum}, and non-classical states of light have become integral to large-scale sensing platforms such as LIGO ~\cite{tse2019quantum}. In many cases, the creation and use of entangled states introduce technical trade-offs that must be balanced against overall sensor performance. Very recently, however, several experiments have begun to demonstrate genuine net gains—achieving performance approaching that of the best sensors operating without entanglement~\cite{eckner2023realizing,yang2025clock,dietze2025entanglement}. Optical atomic clocks have played a prominent role in this progress, benefiting from the overlapping technical requirements of high-precision timekeeping—such as long coherence times, large atom numbers, and precise control—and the implementation of entangling protocols. In the following, we outline several of these experimental advances and discuss near-term proposals that hold promise for broad classes of quantum sensors.

\subsection{Entanglement-enhanced optical clocks operating below SQL} 
Atomic clocks based on optical transitions in neutral atoms and ions are the highest performing devices for time-and-frequency metrology to date, with fractional uncertainties at the $10^{-18}$-level and below~\cite{ludlow2015optical, brewer2019al+, aeppli2024clock, marshall2025high}. As discussed in Sect.~\ref{sec:clocks_theory}, they operate by comparing an oscillator to the frequency of an atomic transition. As illustrated in Eq.~\eqref{eq:instability}, the fractional frequency stability improves with oscillator frequency $\omega_0$, as this frequency determines the number of cycles  counted per second during an interrogation, i.e. by analogy with measuring lengths, it is the density of ticks on this temporal ruler. Meanwhile, $\Delta^2_\phi$ denotes the average phase estimator variance, as defined in Eq.~\eqref{eq:AverageEstVar}, which can be reduced through the use of entangled states and optimized measurement strategies.

\begin{figure*}[t] 
   \centering
   \includegraphics[width=\textwidth]{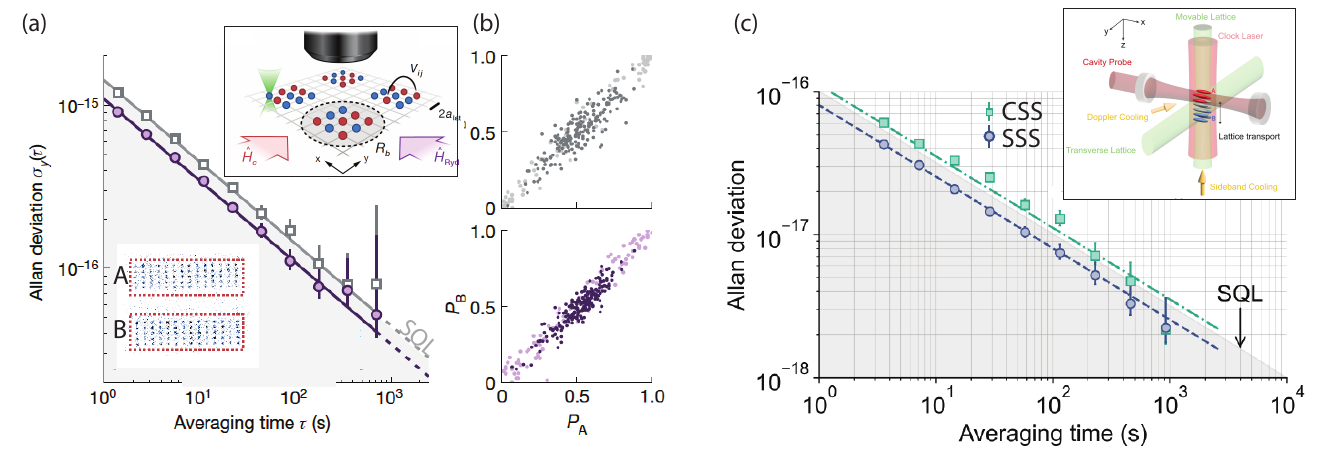}
   \caption{Direct observation of neutral atom optical clocks operating below SQL (a) Rydberg-mediated interactions are used to generate a spin-squeezed optical clock with up to 70 atoms. Two copies of these states are compared in an atom–atom frequency comparison, demonstrating performance below the SQL . (b) Distribution of outcomes showing the excited-state probability of ensemble $A$ plotted against that of ensemble $B$. The reduced noise in the joint distribution provides direct visual evidence of entanglement. Adapted from Ref.~\cite{eckner2023realizing}. (c) Atomic ensembles stored in a conveyor-belt optical lattice are transported into and out of a cavity, where squeezing operations are performed. An atom–atom frequency comparison between the ensembles demonstrates sub-SQL performance with precision at the $10^{-18}$-level. Adapted from Ref.~\cite{yang2025clock}).
    }
   \label{fig:entangled_clocks}
\end{figure*}

Inspired by foundational insights and realizations in atomic clocks based on microwave transitions~\cite{leroux2010implementation,riedel2010atom, chen2011conditional, haas2014entangled, bohnet2014reduced, cox2016deterministic, hosten2016measurement}, a key motivating question in the community has been whether rapid advances in entanglement generation within neutral-atom and ion platforms might be incorporated into optical atomic clocks~\cite{colombo2022entanglement, ye2024essay}. As detailed below, the main optical clock architectures pursuing entanglement-enhanced operation are optical lattice clocks coupled to optical cavities, trapped-ion optical clocks, and neutral-atom clocks that exploit Rydberg-mediated interactions. ~\cite{pedrozo2020entanglement, marciniak2022optimal, eckner2023realizing, robinson2024direct, yang2025clock, dietze2025entanglement}.

\subsubsection{Neutral-atom optical clocks operating below the standard quantum limit} 

\begin{figure*}[t] 
   \centering
   \includegraphics[width=0.8\textwidth]{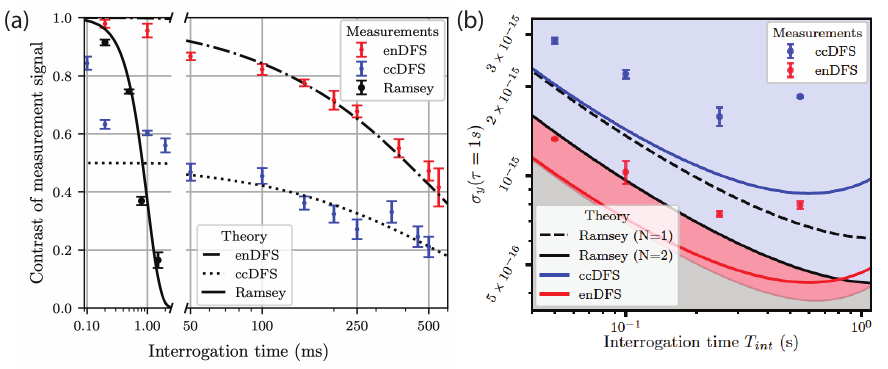}
   \caption{Realizations of entangled optical clocks based on trapped ions operating at the state of the art. (a) Comparison of Ramsey coherence times for a product state of two ions (black), a classically correlated mixed state with reduced average sensitivity to magnetic fields (blue), and an entangled state with similarly suppressed magnetic-field sensitivity (red). (b) Allan deviation at 1-second averaging for different Ramsey interrogation times. The entangled state outperforms the classically correlated state and achieves performance within a factor of two of the current state of the art for trapped-ion clocks~\cite{marshall2025high}. Adapted from Ref.~\cite{dietze2025entanglement}}
   \label{fig:ion_entangled_clocks}
\end{figure*}

Here we review recent progress in the use of entanglement in neutral-atom optical clocks; we also point the reader to another recent review on this topic~\cite{colombo2022entanglement}. In a pioneering demonstration, the first optical atomic clock incorporating entanglement in the form of spin squeezing was reported~\cite{pedrozo2020entanglement}. This result employed collective cavity mediated interactions within a neutral-atom ensemble to generate squeezing~\cite{leroux2010implementation, chen2011conditional, haas2014entangled, bohnet2014reduced, hosten2016measurement}, and showed performance below SQL after subtracting local-oscillator noise. Meanwhile, a tweezer-controlled optical lattice experiment demonstrated the first realization of Bell states in a neutral-atom optical clock, employing Rydberg-mediated interactions and achieving atomic coherence times of several seconds~\cite{schine2022long}. Shortly thereafter, the same system demonstrated quantum-enhanced clock performance below the SQL (see Fig.\ref{fig:entangled_clocks}a) using Rydberg dressing~\cite{zeiher2017coherent}, achieving precision at the $10^{-17}$ level via direct atom–atom frequency comparisons between ensembles of 70 atoms, which reflected the bare performance of the quantum-enhanced sensor~\cite{eckner2023realizing}. Comparable quantum-enhanced precision was achieved using cavity-mediated interactions in an optical lattice clock with $10^4$ atoms, in a setup where  the optical lattices can be transported in and out of the cavity, enabling the interrogation of ensembles containing up to $10^4$ atoms~\cite{robinson2024direct}. Most recently, this cavity-based architecture demonstrated precision reaching the $10^{-18}$ level in direct atom-atom comparisons, which is below the SQL (see Fig.\ref{fig:entangled_clocks}c)~\cite{yang2025clock}. These results establish the compatibility of entangled states with state-of-the-art optical clock performance, showing that entanglement-enhanced clocks can achieve precisions rivaling the accuracies of the leading systems~\cite{brewer2019al+, aeppli2024clock}.

\subsubsection{Entangled trapped-ion clocks}
\label{sec:ionclocks}
Trapped ion experiments have demonstrated entanglement enhanced Ramsey interferometry on optical transitions. These advances build on mature techniques for implementing high-fidelity two-qubit and collective gate operations on optical qubits~\cite{benhelm2008towards, monz201114, akerman2015universal, clark2021high}, enabling, for example, the generation of GHZ states with up to 24 ions~\cite{pogorelov2021compact}, the realization of Heisenberg-limited spectroscopy~\cite{shaniv2018toward}, and the application of variational methods for phase estimation~\cite{marciniak2022optimal}. In practice, the number of ions that can be precisely controlled in a single trap is limited by the constraints of the trapping architecture. Within this context, quantum metrology offers a particularly valuable strategy to enhance clock performance. 

Most recently, an entanglement-enhanced trapped-ion clock was demonstrated (see Fig.\ref{fig:ion_entangled_clocks})~\cite{dietze2025entanglement}. This experiment employed two-ion Bell states to achieve two key improvements. First, by encoding the Bell states such that the ground and excited components of the superposition had identical first-order magnetic-field sensitivities, the system became robust against magnetic noise (see Fig.\ref{fig:ion_entangled_clocks}a). This encoding extended the coherence time by more than two orders of magnitude, with the resulting contrast limited only by the natural lifetime of the clock states. Second, a laser stabilized to these Bell states showed enhanced frequency stability—outperforming a comparable product-state reference—for dark times up to 0.5 seconds (see Fig.\ref{fig:ion_entangled_clocks}b). Together, these advances enabled an ion stability of $7\times10^{-16}$ at 1 second, which is a record for calcium and approaching the record with aluminum~\cite{marshall2025high}. 

\subsubsection{Key challenges}
\label{sec:key_challanges}

Optical clocks enhanced with entangled states face a number of challenges. When stabilizing a local oscillator  to an atomic ensemble, two primary noise mechanisms can limit clock performance. First, phase diffusion becomes problematic when the interrogation time $T$ approaches or exceeds the coherence time $T_{\rm C}$ of the local oscillator, potentially driving the accumulated phase outside the dynamic range of the quantum sensor ~\cite{andre2004stability,leroux2017line}. Second, the Dick effect arises from the presence of dead time between successive interrogations—such as during atomic state preparation, readout, or repumping—when the local oscillator is not referenced to the atoms~\cite{dick1989local}. During these intervals, the local oscillator continues to evolve, but its frequency changes are neither monitored nor corrected, resulting in an aliasing of high-frequency local oscillator noise and thereby degrading the clock’s short-term stability.

Ref.~\cite{schulte2020prospects} presents a detailed analysis of these effects when using squeezed states, showing that they collectively impose a fundamental stability floor, independent of atom number, beyond which further reduction of quantum projection noise offers no improvement. In this regime, spin squeezing can still reduce the number of atoms required to reach this floor, but cannot surpass it. Importantly, this limitation can be mitigated in squeezed-state clocks operating with negligible dead time~\cite{biedermann2013zero, al2015noise}, or in architectures that perform frequency comparisons between two atomic ensembles—an approach intrinsically insensitive to the Dick effect~\cite{hume2016probing, marti2018imaging, young2020half, eckner2023realizing, robinson2024direct, yang2025clock}. Moreover, in sensing applications where signals must be resolved on timescales shorter than the laser coherence time, spin squeezing and increased atom number continue to offer sensitivity enhancements.

While several recent experiments have demonstrated optical atomic clocks with entanglement-enhanced short-term stability, a comprehensive accuracy evaluation has not yet been conducted in any system employing entangled states. A natural question arises as to whether the infrastructure and control operations required to generate entanglement might introduce systematic shifts that compromise overall clock accuracy. This remains an active area of investigation. In many implementations, the entangling operations are performed prior to the interferometric phase accumulation, which is the period most sensitive to transition-shifting perturbations. 

\subsubsection{Opportunities for entanglement in  Optical Clocks}
As discussed in Sect.~\ref{sec:atomicCoherenceTime}, when comparing nearly static frequencies, entanglement does not provide a fundamental advantage. In this regime, optimal clock stability is achieved by operating at the longest possible interrogation time and using (unentangled) coherent states, which is ultimately limited by the atomic coherence time $T_{\rm A}$. 

However, in scenarios where a stable atomic reference is used to detect a rapidly varying signal—i.e., one with characteristic timescales shorter than $T_{\rm A}$—short interrogation times are required to resolve the fast dynamics. In this regime, where decoherence is negligible, entangled states can offer a sensitivity advantage over unentangled sensors. This advantage can be interpreted as an entanglement-enhanced extension of the sensor’s effective bandwidth (see also Ref.~\cite{colombo2022entanglement}).

From this perspective, we identify several practical contexts in which entangled sensors may offer meaningful performance gains. For example, in optical atomic clocks, entangled states could be leveraged to suppress high-bandwidth fluctuations of the local oscillator. As entangling operations continue to improve in robustness and fidelity, such capabilities may be particularly advantageous in transportable optical clocks~\cite{grotti2018geodesy, takamoto2020test, takamoto2022perspective}, where acoustic noise in the $0.01$–$1\mathrm{kHz}$ range can be significant.

Furthermore, optical atomic clocks have recently been proposed as detectors for gravitational waves~\cite{kolkowitz2016gravitational}, whose characteristic oscillation period may coincide or be below the atomic clock coherence time~\cite{abbott2016observation, bailes2021gravitational}. In this regime, entanglement-enhanced clocks could play a role in improving sensitivity. Similarly, optical clocks are already being employed in searches for temporal variation of fundamental constants. Entangled sensors could improve sensitivity within targeted frequency bands, thereby tightening constraints on such variations.

In particular, oscillations in fundamental constants—manifesting as shifts in atomic transition frequencies—can be used to probe for axion-like particles over specific energy ranges. The corresponding frequency windows for these searches can align with those accessible via entanglement-enhanced protocols~\cite{stadnik2015can, safronova2018search, kim2024oscillations}.

Finally, we wish to highlight areas where, at present, we do \emph{not} see obvious applications of entangled states in clocks and spectroscopy. Many experiments aim to achieve ever-increasing precision in the measurement of essentially static quantities. Examples include searches for the electron’s electric dipole moment using Ramsey spectroscopy in molecules, measurements of nuclear quadrupole or higher-order multipole moments via hyperfine spectroscopy, and determinations of static perturbations such as blackbody radiation shifts or Stark/Zeeman shifts in clock states~\cite{safronova2018search}. In such cases, whenever the atomic coherence time $T_{\rm A}$ is finite, the optimal stability attainable with unentangled ensembles is, up to constant factors, the same as that achievable with entangled states, albeit at a different interrogation time. At present, we are not aware of methods that would enable scalable improvements in precision using entangled states for these static applications.

\subsection{Enhancing robustness of entangled sensors }
\label{sec:robustness}
As discussed in Sect.~\ref{sec:frequencyEstimation}, increased entanglement—and the corresponding enhancement in sensitivity—is often accompanied by greater susceptibility to noise, arising from both technical sources (e.g., laser phase noise, frequency inhomogeneities) and fundamental processes such as spontaneous emission. To address these challenges, ongoing research is exploring how sensor performance can be improved through the integration of error mitigation and detection techniques, including quantum error correction and adaptive feedback protocols inspired by advances in quantum information science. Additionally, efforts are directed toward the development of optimized measurement strategies that are more resilient to noise, with the goal of preserving metrological advantage under realistic experimental conditions.

\subsubsection{Error mitigation} 

A prime example of an error mitigation strategy in quantum metrology involves the use of GHZ states. These states are, in principle, optimal for parameter estimation, as they maximize the QFI among all possible quantum states. This advantage stems from their $N$-fold enhanced sensitivity to the signal, relative to unentangled atoms. However, this enhanced susceptibility to signal also renders GHZ states highly vulnerable to various sources of noise.

One notable example of error mitigation using GHZ states is the GHZ cascade protocol, discussed in Sect.~\ref{sec:GHZ_Cascade}. The core challenge addressed by this scheme arises from phase diffusion induced by noise in the reference frequency, which causes the accumulated phase to spread beyond the $[-\pi/(2N), +\pi/(2N)]$ interval—within which an $N$-atom GHZ state can unambiguously resolve the phase. By employing GHZ states of varying sizes, this ambiguity can be mitigated: smaller GHZ states provide coarse, unambiguous estimates of the phase interval, while larger GHZ states contribute enhanced precision in refining the phase estimate within the identified region.

Nonetheless, recent experimental demonstrations of such protocols~\cite{cao2024multi, finkelstein2024universal} emphasize that the practical sensitivity of GHZ states is significantly limited by imperfections in state preparation and environmental noise, particularly magnetic field fluctuations. This inherent fragility has led to ongoing debate regarding the practical utility of GHZ states in metrological settings. These challenges have motivated the exploration of alternative entangled states and measurement strategies that preserve enhanced sensitivity while mitigating susceptibility to external noise.

As discussed in Sect.~\ref{sec:ionclocks}, a recent  quantum-enhanced trapped-ion clock, a Bell state—formally equivalent to a two-qubit GHZ state under local rotations—was employed to cancel first-order magnetic field sensitivity while maintaining high precision ~\cite{dietze2025entanglement}. However, even at the optimal interrogation time, the interferometric sensitivity was ultimately constrained by the finite lifetime of the excited state due to spontaneous emission.

Spontaneous emission errors can be mitigated to a certain extent in GHZ-states sensors, as demonstrated in Ref.~\cite{kielinski2024ghz}. The proposed strategy involves applying the decoding unitary introduced in Sect.~\ref{sec:GHZ_state} after the phase-encoding interrogation time, followed by a projective measurement in the $z$-basis, rather than performing a standard parity measurement. If no spontaneous emission event occurs, the final state remains a coherent superposition of $\ket{\uparrow}^{\otimes N}$ and $\ket{\downarrow}^{\otimes N}$, with relative amplitudes determined by the phase to be estimated.
In contrast, any spontaneous emission event populates orthogonal states, thereby making the error detectable. This enables a simple form of post-selection, where all measurement outcomes that do not result in all the atoms being measured in the ground or excited state.

This error-detection scheme effectively suppresses the impact of spontaneous emission—provided that the probability of such events remains low. As the system size $N$ increases, the likelihood of an error occurring during interrogation grows, eventually requiring a reduction in the interrogation time to maintain a manageable error rate. Consequently, while this strategy offers a clear advantage for moderate system sizes, asymptotically in $N$ it yields only a constant-factor improvement over the SQL. Nonetheless, for system sizes up to approximately 80 atoms, it represents the most effective measurement strategy in the presence of spontaneous emission.

Another theoretical approach, aimed more directly at error suppression rather than mitigation, involves the engineering of superatoms composed of two fermionic atoms confined to a single lattice site. By carefully designing the dipolar interactions and leveraging Pauli blocking, spontaneous emission can be significantly suppressed, effectively realizing a decoherence-free superatom~\cite{pineiro2019dark}. This concept exploits the interplay between the structured dipolar interactions and the fermionic exchange symmetry, allowing the system to occupy collective dark states in which spontaneous emission is strongly suppressed. As a result, long-lived entangled states with greatly reduced radiative decay can be created, offering a promising route toward robust quantum sensing and metrology.

\subsubsection{Quantum error correction for metrology}
\label{sec:QEC}

A natural question that arises in the context of quantum sensing is whether quantum error correction (QEC) can be employed to enhance sensor performance. Specifically, one may ask whether QEC can be used to correct spontaneous emission errors, thereby overcoming limitations set by the atomic coherence time in optical atomic clocks. Such an approach could potentially enable coherent interrogation over durations approaching the total averaging time, thereby preserving a favorable Allan deviation scaling of $\sigma_y \propto 1/\tau$ at long times.

However, a no-go theorem has established that QEC does not offer an advantage for many common noise mechanisms, including spontaneous emission, local dephasing, and global dephasing, depending on the signal being measured ~\cite{demkowicz2017adaptive, zhou2018achieving}. The underlying intuition~\cite{kessler2014quantum} is that only noise processes that act in a manner distinct to the signal encoding can be corrected without simultaneously erasing the signal. Consequently, it becomes clear why neither local nor global dephasing—both of which act through the same operators as the signal—can be mitigated using QEC.

Moreover, spontaneous emission errors belong to a class of noise processes for which QEC does not enable performance beyond the SQL. This is the case when the signal Hamiltonian lies within the Lindblad span of the noise operators. More precisely, if the signal-generating operator $\sigma_z^{(k)}$, can be expressed as a linear combination of the set $\{ L_k, L_k^{\dagger}, L_k^{\dagger} L_{k'},\mathds{1}\}$, where ${L_k}$ are the Lindblad jump operators characterizing the noise, then the Hamiltonian is said to be in the Lindblad span. When this condition is satisfied, the optimal achievable Allan deviation scales as $\sigma_y \propto 1/\sqrt{N T \tau}$, with a prefactor that depends on the detailed structure of the noise via the specific form of the jump operators.

By contrast, if the noise is characterized by jump operators like $\sigma_x$, QEC is indeed possible. In these cases, it is theoretically feasible to extend the interrogation time indefinitely, or until limited by other uncorrected noise sources.

An alternative scenario in which QEC can enhance sensing performance arises when the spatial correlations of the noise differ from those of the signal ~\cite{layden2018spatial}. Even when the noise and signal couple through the same local operators (e.g., dephasing via $\sigma_z$ operators), their distinct spatial structure can be exploited for error correction. This situation is similar to dynamical decoupling, where spatial or temporal filtering enables discrimination between signal and noise.

\subsubsection{Overcoming finite detection fidelity}

A general challenge for large-scale quantum sensors is that achieving sensitivity at the SQL already requires high detection fidelity; insufficient detection performance can obscure SQL-level operation entirely. The impact of finite detection fidelity becomes even more severe when the initial probe state is entangled, as measurement errors can rapidly degrade or mask the metrological advantage. This limitation has been a persistent obstacle to demonstrating entanglement in large quantum systems, particularly in cavity-based platforms where atoms are typically measured collectively via the cavity mode rather than through single-site–resolved detection, or in systems with untrapped atoms where spatially resolved measurements are not feasible.

One strategy to overcome detection limitations is the use of time-reversal protocols, in which the entangling unitary applied before phase encoding is inverted after the phase is imprinted ~\cite{davis2016approaching, frowis2016detecting, macri2016loschmidt, nolan2017optimal, haine2018using, colombo2022time, liu2022nonlinear, volkoff2022asymptotic, li2023generalized, liu2023cyclic}, or, more generally, the use of decoding unitaries to measure in an entangled basis. This concept was first developed for SSSs generated via OAT~\cite{davis2016approaching}. In this setting, the phase encoding corresponds to a rotation of the squeezed state. If this rotation is sufficiently small, reversing the OAT dynamics returns the system to a state closely resembling the initial CSS, but now rotated by an angle amplified in proportion to the amount of metrologically useful entanglement produced in the initial quench. This amplification enables Heisenberg scaling even without single-atom–resolved number detection~\cite{davis2016approaching}.

Using such a protocol, Heisenberg scaling was experimentally realized in a cavity-QED system using nuclear-spin qubits of $^{171}$Yb~\cite{colombo2022time}, reaching up to 11.8 dB of enhanced phase resolution with $\sim 300$ atoms. As previously demonstrated, the gain realized with the nuclear spin can also be mapped to the optical clock~\cite{pedrozo2020entanglement}. In general, this and related protocols have been explored in other cavity-QED systems~\cite{hosten2016quantum}, in force-sensing applications with two-dimensional ions~\cite{gilmore2021quantum}, as well as extensions of these time-reversal protocols to Hamiltonians featuring chaotic dynamics~\cite{oszmaniec2016random,li2023improving}. 

\subsection{Scalable squeezing in short-range interacting systems}
\label{sec:short_range_squeezing}

\begin{figure*}[t] 
   \centering
   \includegraphics[width=\textwidth]
   {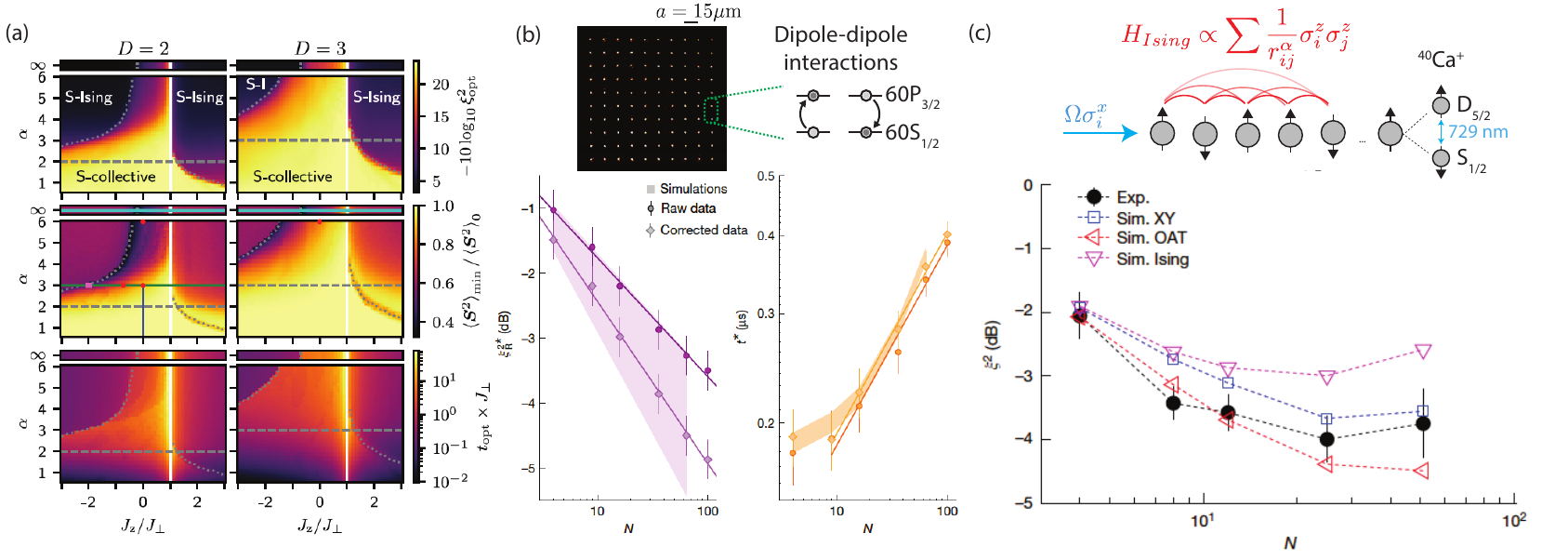}
   \caption{Spin squeezing for power-law interactions with exponent $\alpha$ and varying anisotropy ratios in the Heisenberg model of Eq.\eqref{eq:XXZ}, where we use $\chi, \chi’$ in place of $J_{\rm Z}, J_{\perp}$. The squeezing, total angular momentum, and time to generate maximal squeezing are calculated in two (D=2) and three (D=3) dimensions. Regions of scalable squeezing are observed despite the short-range nature of the interactions (adapted from Ref.~\cite{perlin2020spin}). (b) Experimental realization using a tweezer array of ${}^{87}$Rb atoms (top left), interacting with Rydberg-mediated dipolar flip-flop interactions (top right). Scalable squeezing is observed for the accessible system sizes (bottom left), with the time ($t^*$) to reach optimal squeezing increasing with system size due to the finite interaction range (bottom right) (adapted from Ref.~\cite{bornet2023scalable}). (c) Enhanced squeezing beyond the pure Ising limit realized in a one-dimensional chain of Ca$^+$ ions. Application of a strong transverse field maps the Ising model to an effective XXZ model with anisotropy $\chi’=2\chi$, enabling propagation of correlations across the system and thereby improved squeezing performance. Adapted from Ref.~\cite{franke2023quantum}.
   }
   \label{fig:short_range}
\end{figure*}

Another key question is to determine the conditions under which many-body dynamics can generate metrologically useful entanglement that is scalable. By scalable, we mean that the phase estimation variance scales as $\Delta_{\phi}^2 \propto N^{-\alpha}$ with $\alpha > 1$, corresponding to an improvement beyond the SQL. This is fundamentally a question of many-body physics: how do quantum correlations arising in an interacting system give rise to entanglement that can be harnessed for enhanced metrology?

A key implication is that scalable enhancement requires multipartite entanglement that spans the entire ensemble, as the QFI is a measure of the degree of multipartite entanglement (see Sect.~\ref{sec:qfi_limits}). The paradigmatic model for dynamically generating such entanglement is the OAT Hamiltonian Eq.~\eqref{eq:H_OAT} in which each spin interacts with every other spin equally. This all-to-all coupling results in entanglement that extends across the full system and enables a spin-squeezing parameter that improves with system size. Realizations of this Hamiltonian have been achieved in various platforms, including trapped ions ~\cite{leibfried2004toward, benhelm2008towards}, optical cavities ~\cite{li2022collective, greve2022entanglement}, Bose-Einstein condensates via elastic collisions ~\cite{esteve2008squeezing, riedel2010atom}, as well as more recently in Rydberg tweezer arrays, albeit so far only for system sizes of up to 9 atoms~\cite{cao2024multi}.

In contrast, many more physical systems are characterized by finite-range interactions, prompting the question of whether and under what conditions such systems can support scalable squeezing. A limiting case is provided by nearest-neighbor Ising interactions, described by
\begin{equation}
    H_{\rm I} = \chi \sum_{\langle k,l \rangle} \sigma_z^{(k)} \sigma_z^{(l)},
\end{equation}
where entanglement develops only between directly interacting spins, and quantum correlations do not propagate beyond the interaction range due to the absence of a light cone ~\cite{foss2013nonequilibrium}. As a consequence, no scalable spin squeezing can be achieved in such systems. Even in systems with slightly longer-range interactions—such as those realized in recent experiments ~\cite{eckner2023realizing}—the amount of squeezing exceeds nearest neighbor Ising interactions but is still not scalable ~\cite{gil2014spin}.

More generally, for Ising-type models with power-law decaying interactions,
\begin{equation}
    H_{{\rm I},\alpha} = \chi \sum_{k,l} \frac{1}{|\bm{r}_k - \bm{r}_l|^\alpha} \sigma_z^{(k)} \sigma_z^{(l)},
\end{equation}
scalable spin squeezing is only possible if the decay exponent satisfies $\alpha < D$, where $D$ is the dimensionality of the system ~\cite{foss2016entanglement}. Here, $\bm{r}_k$ denotes the spatial position of spin $k$.

This behavior contrasts with models based on anisotropic Heisenberg interactions, such as
\begin{equation}
\label{eq:XXZ}
H_{\rm XXZ} = \sum_{k,l} \frac{\chi \, \bm{\sigma}^{(k)} \cdot \bm{\sigma}^{(l)} + (\chi-\chi’ )\, \sigma_z^{(k)} \sigma_z^{(l)}}{|\bm{r}_k - \bm{r}_l|^{\alpha}},
\end{equation}
where $\bm{\sigma} = (\sigma_x, \sigma_y, \sigma_z)$. These models are capable of generating longer-range quantum correlations, and recent theoretical work (see Fig.~\ref{fig:short_range}a) has explored their potential for generating scalable squeezing ~\cite{frerot2017entanglement, perlin2020spin, comparin2022scalable, young2023enhancing}. 

Interestingly, it has been shown that scalable spin squeezing can still be achieved even when the interaction range satisfies $\alpha > D$, provided the system operates close to the Heisenberg point $\chi’ \approx \chi$ ~\cite{perlin2020spin}. This advantage, however, comes at the cost of significantly longer timescales for entanglement generation (see Fig.\ref{fig:short_range}b), which is detrimental in the presence of decoherence during state preparation.

The underlying mechanism can be understood in the perturbative regime $\chi \gg |\chi’|$, where the dominant Heisenberg interaction energetically favors spin alignment. In this limit, permutation-symmetric states are separated from states with broken permutation symmetry by a large energy gap that cannot be bridged by the anisotropic perturbation. Consequently, the dynamics can be effectively projected into the permutation-symmetric subspace. Within this subspace, the residual Ising interaction acts as an effective infinite-range OAT Hamiltonian, enabling scalable squeezing despite the finite interaction range—though with an effective interaction strength that decreases as the anisotropy is reduced.

Motivated by these theoretical developments and the desire to enhance the amount of spin squeezing, recent experiments have begun to explore these ideas ~\cite{franke2023quantum, bornet2023scalable} (see Fig.~\ref{fig:short_range}b,c). In a one-dimensional ($D=1$) trapped-ion array, enhanced spin squeezing and the generation of long-range correlations were demonstrated by introducing a strong transverse field to a system governed by a power-law Ising interaction with exponent $\alpha = 1$~\cite{franke2023quantum}. This configuration effectively realizes a spin-exchange Hamiltonian, which corresponds to the XXZ Hamiltonian with anisotropy parameter $\chi’ = 2\chi$, leading to a significant improvement in squeezing relative to the native Ising interaction alone. Recent Rydberg-dressing-based squeezing experiments - in optical clocks~\cite{eckner2023realizing} and also in atomic ensembles of microwave qubits~\cite{hines2023spin} -  could benefit from the application of a strong-transverse  to improve the scaling of the squeezing \cite{young2023enhancing}, as it has been previously demonstrated in trapped ions~\cite{franke2023quantum} and atomic gases~\cite{borish2020transverse}. Recent theoretical work has further proposed a general framework linking the presence of scalable squeezing in the dynamics of a given Hamiltonian $H$ to the existence of long-range order at finite temperature in the same model ~\cite{block2024scalable}.

Without enhancing the performance of an underlying Ising interaction, spin squeezing was achieved in a two-dimensional Rydberg atom array, where the qubit states correspond to opposite-parity Rydberg levels (Fig.~\ref{fig:short_range}b). In this system, the atoms interact via dipolar spin-exchange interactions characterized by a power-law decay with exponent $\alpha = 3$, enabling the generation of metrologically-useful entanglement that improves with atom number~\cite{bornet2023scalable}.

An alternative direction, explored at the theoretical level, is the use of finite-range interactions within variational quantum circuits that are optimized to maximize the achievable spin squeezing ~\cite{kaubruegger2019variational}. While this approach does not fundamentally overcome the limitations on scalable squeezing imposed by the nature of the native interactions, it offers a practical route to optimally harness available resources on a given quantum sensing platform.

\section{Outlook}

Here, we provide a broader perspective on several emerging directions in quantum metrology. For a more comprehensive discussion of quantum metrological applications on the horizon, we refer the reader to Ref.~\cite{ye2024essay}.

The fundamental limits of single-parameter quantum metrology are by now well established. At present, however, the most pressing open questions concern how to approach these limits as efficiently as possible with the quantum sensors that are currently available, thereby closing the gap between theoretically predicted advantages and what can be achieved in practice with experimentally developed devices. A central challenge in this respect is to devise hardware-efficient strategies for achieving metrological enhancement. This involves identifying practical methods to prepare entangled states and to realize the optimized measurements using the restricted capabilities of existing sensing platforms, which are typically far from universal quantum computers.

Several promising avenues are beginning to emerge. One direction is the use of adaptive measurement strategies: while such approaches do not surpass the performance of fully optimal generic measurements ~\cite{kurdzialek2023using}, they can nevertheless provide practical advantages by simplifying the realization of near-optimal measurements with the hardware available in experiments ~\cite{direkci2024heisenberg}. Another direction is the adaptation of variational quantum circuits to the native resource operations offered by specific sensing platforms ~\cite{kaubruegger2019variational,kaubruegger2021quantum}. This makes it possible to perform on-device optimization directly in the presence of realistic noise sources ~\cite{marciniak2022optimal}, which are often difficult to fully characterize or simulate. Such hardware-adapted and noise-aware strategies are therefore particularly promising for translating the abstract limits of quantum metrology into experimentally relevant gains.

Another important direction in contemporary quantum metrology is the use of sensor networks to estimate functions of the phases encoded across multiple nodes \cite{zhang2021distributed}. Depending on the spatial arrangement of the sensors, these functions may correspond to the mean phase, spatial gradients, or even higher-order curvature. Such capabilities extend the scope of quantum metrology to the characterization of spatially varying fields.

Illustrative application of such sensor networks is provided by optical clock networks. When distributed over large spatial separations, these networks can function as detectors of gravitational waves by exploiting the differential phase shifts imprinted between distant nodes ~\cite{kolkowitz2016gravitational}. At the opposite extreme, networks of closely spaced clocks have recently been employed to resolve gravitational redshifts on the scale of the spatial extent of the trapped atomic ensembles themselves, thereby enabling precision tests of gravity at microscopic length scales ~\cite{bothwell2022resolving, zheng2022differential}.

Furthermore, sensor networks also provide a natural route to suppressing spatially correlated noise~\cite{layden2018spatial}. In particular, when the spatial correlations of the noise differ sufficiently from the spatial structure of the signal to be estimated, one can design measurement strategies that preserve sensitivity to the desired signal while remaining robust against noise components that are effectively orthogonal to the signal correlations~\cite{altenburg2017estimation, eldredge2018optimal, sekatski2020optimal, hamann2022approximate, hainzer2024correlation, kaubruegger2025lieb, mamaev2025non, chu2025reconfigurable}. Moreover, sensor networks enable efficient reconstruction of spatially varying fields by treating the encoded phases as samples of an underlying signal and applying compressed sensing techniques to recover the full profile from limited measurements~\cite{baamara2023quantum}.

Another direction in contemporary quantum metrology, which we have thus far not addressed, is the simultaneous estimation of multiple parameters ~\cite{liu2020quantum,demkowicz2020multi}, where the parameters are encoded by processes that do not commute with one another. Vector field sensing provides an example of such a multiparameter estimation problem ~\cite{baumgratz2016quantum,gorecki2022multiparameter,kaubruegger2023optimal,vasilyev2024optimal}. Multiparameter estimation introduces an additional layers of complexity. One is associated with measurement incompatibility. This arises when the optimal measurements for different parameters do not commute and therefore cannot be implemented simultaneously on the same quantum system. As a result, the quantum Cramér–Rao bound cannot, in general, be saturated for all parameters simultaneously. Moreover, the notion of dynamic range becomes more subtle than in the single-parameter case, particularly because adaptive strategies that continually shift the sensor to its optimal operating point are hampered by the noncommutativity of the parameter-encoding processes. These challenges underscore both the richer structure and the increased practical complexity of multiparameter quantum metrology in contrast to single-parameter estimation.

On the experimental front, entangled optical clocks are now achieving precisions at the $10^{-17}$ to $10^{-18}$ level~\cite{eckner2023realizing,robinson2024direct,yang2025clock}, making the near-term prospects for their operation at the time-keeping frontier increasingly promising. Further improvements are still required to match the stabilities of the best unentangled clocks, which have already reached $10^{-18}$ stability at 1 second and atom–atom comparison precisions at the $10^{-21}$ level~\cite{bothwell2022resolving,zheng2022differential}. With realistic advances in system size and coherence times, entangled neutral-atom clocks are expected to become competitive with unentangled devices for resolving effects that require short interrogation times. Beyond their metrological potential, such systems may also provide a unique test bed for probing the interface between gravity and quantum mechanics~\cite{pikovski2015universal,roura2020gravitational}.

While this review has primarily focused on clocks as a representative platform for quantum sensing, many existing and emerging sensors are now exploiting entangled states. Atom interferometers based on spatial superpositions of massive particles enable precision measurements of gravity and fundamental constants, as well as stringent tests of the equivalence principle, general relativity, and possible new forces~\cite{peters1999measurement,muller2010precision,rosi2014precision,zhou2015test,jaffe2017testing,parker2018measurement,asenbaum2020atom,panda2024measuring}. A particularly notable recent milestone is the first demonstration of an entangled atom interferometer, which reported sensitivity 2.5 dB below the SQL~\cite{greve2022entanglement}. 

In the context of force sensing, the quantized motion of trapped ions provides a powerful platform for displacement detection. Owing to the ionic charge, such protocols are naturally applicable to the measurement of both the amplitude and phase of oscillating electric forces. Quantum-enhanced precision can be achieved by employing highly excited Fock states, where the nonclassical motional character, rather than entanglement, enables sensitivity beyond classical limits.  Importantly, these protocols have been demonstrated in a phase-insensitive regime ~\cite{wolf2019motional}, a key advantage when the force phase is unknown at the time of state preparation, which is particularly relevant for signals with a linewidth exceeding the measurement bandwidth, where adaptive phase matching is not feasible. Along similar lines, superpositions of Fock states have been employed for enhanced sensing of oscillator mode frequencies ~\cite{wolf2019motional,mccormick2019quantum}. Furthermore, in Penning traps with hundreds of ions, two-mode squeezed states of the combined system of spins and motion have been realized for displacement and electric-field sensing, with proposed applications to the detection of dark photons and axions ~\cite{gilmore2021quantum}.

In parallel with advances in atomic platforms, spin ensembles embedded in solid-state materials provide a complementary route for quantum sensing~\cite{schirhagl2014nitrogen,rodgers2021materials,rovny2024nanoscale}. Prominent examples include vacancy centers in diamond, which enable high-spatial-resolution studies of magnetic and electric fields as well as in-situ characterization of material properties such as charge transport~\cite{maze2008nanoscale,dolde2011electric,cooper2014time,liu2019nanoscale,kolkowitz2015probing,jenkins2022imaging, marchiori2022nanoscale}. Their small form factor and room-temperature operability further extend their applicability to diverse environments, including biological settings~\cite{grinolds2013nanoscale,le2013optical,kucsko2013nanometre}. While such solid-state spin systems generally lack the degree of single-particle control available in atomic platforms, the intrinsic interactions between spins—for example, dipolar couplings among nitrogen-vacancy centers—offer a natural mechanism for entanglement generation. To that end, recent experiments have reported exciting progress in spin squeezing of solid-state spin ensembles mediated by dipolar interactions~\cite{wu2025spin}. 

Zooming out, the advent of systems with potentially hundreds to thousands of qubits combining a high-degree of single-qubit control with capabilities in sensing opens the door to a number of directions~\cite{bluvstein2024logical,tao2024high,norcia2024iterative, cao2024multi, finkelstein2024universal}. Importantly, these are systems are potentially capable of realizing deep quantum circuits for the preparation and decoding of quantum states used for parameter estimation.  We anticipate that this synthesis of quantum computing and quantum metrology will motivate new directions catalyzed by progress in each of these areas. Theoretically, the anticipation of such systems has already resulted in a significant body of research into the use of quantum error correction to metrology, of which this review only scratches the surface. Furthermore, the prospect of performing quantum processing directly on sensor data has motivated reformulations of sensing tasks to more closely align with computational problems addressed by established quantum algorithms. Examples include Grover-inspired search protocols for detecting weak oscillating fields of unknown frequency~\cite{allen2025quantum}, as well as quantum algorithms for postprocessing optical imaging information~\cite{mokeev2025enhancing}. These are just some examples,  we expect that the versatility of such platforms will spark unforeseen advances, which lie at the intersection of many fields in quantum information science. 

\begin{acknowledgement}
We thank Alec Cao, Nelson Darkwah Oppong, Shimon Kolkowitz, James K. Thompson, Ana Maria Rey, and Diego Fallas Padilla for their careful reading of the manuscript and for providing helpful suggestions, and also Liang Jiang for clarifying discussions. RK acknowledges funding by the German National Academy of Sciences Leopoldina under grant LPDS 2024-08, the NSF JILA-PFC PHY-2317149 and NSF QLCI awards OMA-2016244 and  OMA-2120757, the U.S. Department of Energy, Office of Science, National Quantum Information Science Research Centers, Quantum Systems Accelerator. AMK acknowledges support from the National Institute of Standards and Technology. 

\end{acknowledgement}

\bibliographystyle{spphys}
\bibliography{ref}

\end{document}